\documentclass[conference]{IEEEtran}
\IEEEoverridecommandlockouts

\usepackage[switch]{lineno}

\usepackage{amsmath,amssymb,amsfonts}
\usepackage{algorithmic}
\usepackage{graphicx}
\usepackage{textcomp}
\usepackage{array}
\usepackage{xcolor}
\def\BibTeX{{\rm B\kern-.05em{\sc i\kern-.025em b}\kern-.08em
    T\kern-.1667em\lower.7ex\hbox{E}\kern-.125emX}}

\usepackage[style=numeric]{biblatex}
\usepackage{comment}
\usepackage{csvsimple}
\usepackage{caption}
\usepackage{subcaption}
\usepackage{siunitx}

\makeatletter
\@namedef{ver@lineno.sty}{9999/12/31}
\@namedef{opt@lineno.sty}{}
\makeatother
\usepackage[frozencache]{minted}
\setminted{fontsize=\footnotesize}

\usepackage{microtype}

\usepackage[hidelinks]{hyperref}
\usepackage{url}
\usepackage[nameinlink]{cleveref}
\usepackage{tikz}
\usepackage{environ}
\usepackage{pgfplots}
\usetikzlibrary{fit}
\usetikzlibrary{positioning}
\usetikzlibrary{decorations.pathreplacing,calligraphy}
\pgfplotsset{compat=1.18}

\usepackage
[
   acronyms,%
   shortcuts,%
   nogroupskip%
]{glossaries-extra}

\GlsXtrEnableEntryUnitCounting{acronym}{0}{page}
\setabbreviationstyle[acronym]{long-short}

\newacronym{SSA}{SSA}{Static Single-assignment Form}
\newacronym{IR}{IR}{Intermediate Representation}
\newacronym{DPS}{DPS}{Destination-Passing Style}
\newacronym{DSL}{DSL}{Domain-Specific Language}
\newacronym{API}{API}{Application Programming Interface}
\newacronym{CPU}{CPU}{Central Processing Unit}
\newacronym{GPU}{GPU}{Graphics Processing Unit}
\newacronym{CAD}{CAD}{Computer-Aided Design}
\newacronym{IDL}{IDL}{Idiom Description Language}
\newacronym{BLAS}{BLAS}{Basic Linear Algebra Subprograms}
\newacronym{RAM}{RAM}{Random Access Memory}
\newacronym{LIAR}{LIAR}{Latent Idiom Array Rewriting}
\newacronym{HPC}{HPC}{High-Performance Computing}
\newacronym{SMR}{SMR}{Source Matching and Rewriting}
\newacronym{CFG}{CFG}{Control-Flow Graph}

\renewcommand*{\glslinkcheckfirsthyperhook}{%
  \ifnum\glsentrycurrcount\glslabel>0
   \setkeys{glslink}{hyper=false}%
  \fi
}

\usepackage{xspace}

\usepackage{array}   %
\newcolumntype{L}{>{$}l<{$}} %

\newcommand{\debruijnsubstituteoperator}{\mathbf{subst}}
\newcommand{\debruijnsubstitute}[2]{\debruijnsubstituteoperator\left(#1, #2\right)}

\newcommand{\debruijnlambda}[1]{\ensuremath{\lambda}\ #1}
\newcommand{\debruijnindex}[1]{\texttt{$\bullet_\texttt{#1}$\xspace}}
\newcommand{\labstraction}[0]{$\lambda$-abstraction\xspace}
\newcommand{\lcalculus}[0]{$\lambda$-calculus\xspace}
\newcommand{\lCalculus}[0]{$\lambda$-Calculus\xspace}
\newcommand{\breduction}[0]{$\beta$-reduction\xspace}

\newcommand{\techniquename}{\ac{LIAR}\xspace}
\newcommand{\Techniquename}{\techniquename}

\newcommand{\egraph}{e-graph\xspace}
\newcommand{\egraphs}{e-graphs\xspace}

\newcommand{\eGraphs}{e-Graphs\xspace}
\newcommand{\enode}{e-node\xspace}
\newcommand{\enodes}{e-nodes\xspace}
\newcommand{\eNode}{e-Node\xspace}
\newcommand{\eNodes}{e-Nodes\xspace}
\newcommand{\eclass}{e-class\xspace}
\newcommand{\eclasses}{e-classes\xspace}

\newcommand{\costfunc}{\mathit{cost}\xspace}
\newcommand{\costof}[1]{\mathit{cost}\left(#1\right)}

\newcommand{\benchmarkname}[1]{\textit{#1}}
\newcommand{\apiname}[1]{\textbf{\texttt{#1}}}

\newcommand{\eg}{\textit{e.g.,}\xspace}

\newcommand{\typeconstraints}[1]{}

\newcommand{\gemv}[1]{\ensuremath{\apiname{gemv}^{\texttt{#1}}}}
\newcommand{\gemm}[2]{\ensuremath{\apiname{gemm}^{\texttt{#1,#2}}}}
\newcommand{\irbuild}[2]{\texttt{build~#1~#2}}
\newcommand{\irbuildf}[2]{\irbuild{#1}{(\debruijnlambda{#2})}}

\makeatletter
\newsavebox{\measure@tikzpicture}
\NewEnviron{scaletikzpicturetowidth}[1]{%
  \def\tikz@width{#1}%
  \begin{lrbox}{\measure@tikzpicture}%
  \BODY
  \end{lrbox}%
  \pgfmathparse{#1/\wd\measure@tikzpicture}%
  \BODY
}
\makeatother

\makeatletter
\newbox\tikz@sand@box
\newcount\tikz@scope@depth
\def\scopenode[#1]#2{%
    \begin{pgfinterruptboundingbox}%
        \advance\tikz@scope@depth111\relax%
        \begin{scope}[name=tempscopenodename,at={(0,0)},anchor=center,#1]%
            \global\let\tikz@fig@name\tikz@fig@name%
            \global\let\tikz@node@at\tikz@node@at%
            \global\let\tikz@anchor\tikz@anchor%
        \end{scope}%
        \let\tikz@scopenode@name\tikz@fig@name%
        \let\tikz@scopenode@at\tikz@node@at%
        \let\tikz@scopenode@anchor\tikz@anchor%
        \setbox\tikz@sand@box=\hbox{%
            \begin{scope}[local bounding box=tikz@sand@box\the\tikz@scope@depth,#1]%
                #2%
            \end{scope}%
        }%
        \setbox\tikz@sand@box=\hbox{}%
        \begin{scope}[local bounding box=\tikz@scopenode@name]%
            \pgftransformshift{\pgfpointanchor{tikz@sand@box\the\tikz@scope@depth}{\tikz@scopenode@anchor}%
                               \pgf@x-\pgf@x\pgf@y-\pgf@y}%
            \pgftransformshift{\tikz@scopenode@at}%
            \begin{scope}[#1]%
                #2
            \end{scope}%
        \end{scope}%
        \pgfkeys{/pgf/freeze local bounding box=\tikz@scopenode@name}%
        \global\let\tikz@scopenode@name@smuggle\tikz@scopenode@name%
    \end{pgfinterruptboundingbox}%
    \path(\tikz@scopenode@name@smuggle.south west)(\tikz@scopenode@name@smuggle.north east);%
    \draw[#1](\tikz@scopenode@name@smuggle.south west)rectangle(\tikz@scopenode@name@smuggle.north east);%
}
\makeatother

\DeclareRobustCommand\circled[1]{\tikz[baseline=(char.base)]{
    \node[shape=circle,fill,inner sep=1pt] (char) {\textcolor{white}{\scriptsize #1}};}}

\newcommand{\reviewonly}[1]{}

\definecolor{highlightblue}{rgb}{0.54, 0.81, 0.94}

\usepackage{ifthen}
\makeatletter
\newcounter{IEEE@bibentries}
\renewcommand\IEEEtriggeratref[1]{%
  \renewbibmacro{finentry}{%
    \stepcounter{IEEE@bibentries}%
    \ifthenelse{\equal{\value{IEEE@bibentries}}{#1}}
    {\finentry\@IEEEtriggercmd}
    {\finentry}%
  }%
}
\makeatother

\begin{document}
\reviewonly{\linenumbers}

\title{Latent Idiom Recognition for a Minimalist Functional Array Language using Equality Saturation\\

}

\author{%
\IEEEauthorblockN{Jonathan Van der Cruysse}
\IEEEauthorblockA{%
    \textit{McGill University} \\
    Montreal, Quebec, Canada \\
    jonathan.vandercruysse@mail.mcgill.ca
}
\and
\IEEEauthorblockN{Christophe Dubach}
\IEEEauthorblockA{%
    \textit{McGill University \& Mila} \\
    Montreal, Quebec, Canada \\
    christophe.dubach@mcgill.ca
}
}

\maketitle

\begin{abstract}
Accelerating programs is typically done by recognizing code idioms matching high-performance libraries or hardware interfaces.
However, recognizing such idioms automatically is challenging.
The idiom recognition machinery is difficult to write and requires expert knowledge.
In addition, slight variations in the input program might hide the idiom and defeat the recognizer.

This paper advocates for the use of a minimalist functional array language supporting a small, but expressive, set of operators.
The minimalist design leads to a tiny sets of rewrite rules, which encode the language semantics.
Crucially, the same minimalist language is also used to encode idioms.
This removes the need for hand-crafted analysis passes, or for having to learn a complex domain-specific language to define the idioms.

Coupled with equality saturation, this approach is able to match the core functions from the BLAS and PyTorch libraries on a set of computational kernels.
Compared to reference C kernel implementations, the approach produces a geometric mean speedup of 1.46$\times$ for C programs using BLAS, when generating such programs from the high-level minimalist language.

\end{abstract}


\section{Introduction}

Generating high-performance code for today's heterogeneous specialized hardware is challenging.
A promising approach is to automatically rewrite specific idioms found in programs as highly optimized library calls~\cite{carvalho2021kernelfarer,ginsbach2018automatic}.
This decouples the compiler's pattern recognition from the hardware-specific knowledge embedded in the library implementation.

However, the library is only useful when idioms are found.
Most prior work~\cite{carvalho2021kernelfarer,ginsbach2018automatic,kawahito2013idiom} encodes idioms in the compiler at a low level of abstraction.
Crafting low-level idioms is tedious, requiring expert compiler knowledge and dedicated analysis passes.
Furthermore, recognition might fail in response to minor changes to the input program.
In short, idiom recognition faces two challenges: to specify idioms in a high-level language and to be robust to minor changes to the input program.

To solve the first challenge, this paper proposes to express both programs and idioms using a high-level functional array language.
Functional array programming for high-performance computing has become increasingly popular in recent years~\cite{henriksen17futhark,ragankelley2013halide,steuwer2017lift} %
and the concise, high-level nature of functional languages simplifies code pattern detection and rewriting~\cite{jones2001playing}.

In a functional language, the second challenge of robust pattern detection amounts to finding hidden idioms.
Consider the following vector sum program: $\texttt{sum(v)} = \texttt{fold (+) 0 v}.$
If we have at our disposal a library function that can quickly perform such a sum, then we rewrite the program as a call to that function.
However, if the library supports more general primitives, the rewriting problem becomes more complicated.

Suppose the library has a function \apiname{dot} that performs a dot product and a function \apiname{fill} that creates an array of identical elements.
A human could combine both to implement the vector sum: $\texttt{sum(v)} = \texttt{\apiname{dot}(v, \apiname{fill}(1))}$.
A pattern matcher does not have human intuition and would be hard-pressed to find this solution as neither the idiom corresponding to \apiname{dot} nor that for \apiname{fill} appear in the original program.

This paper proposes to find such intuitive solutions using \techniquename, a trustworthy technique that finds and exploits \emph{latent} idioms using equality saturation.
Equality saturation~\cite{tate2009equality} discovers all possible program variants encoded as a finite data structure by applying rewrite rules until a fixed point is reached.
\techniquename relies on a minimalist \ac{IR} based on a few simple functional programming primitives, resulting in a compact set of rewrite rules.
This makes it easier to capture the essential structure of a program without getting bogged down in language-specific details.
By using equality saturation to apply rewrite rules that capture both the library-independent semantics of the \ac{IR} and library-specific idioms, \techniquename efficiently transforms programs to expose hidden idioms and improve program efficiency.

This paper shows how this technique can be applied on a subset of the \ac{BLAS}~\cite{blackford2002updated} and PyTorch~\cite{paszke2019pytorch} libraries.
To evaluate \techniquename's effectiveness, this work applies it to custom kernels, and to linear algebra and numerical simulation kernels from the PolyBench suite.
The results show that \techniquename leads to significant improvements in program efficiency.
Overall, this work demonstrates that equality saturation is a powerful tool for idiom recognition, and that \techniquename can be easily adapted to different libraries by providing appropriate idiom descriptions for those libraries.

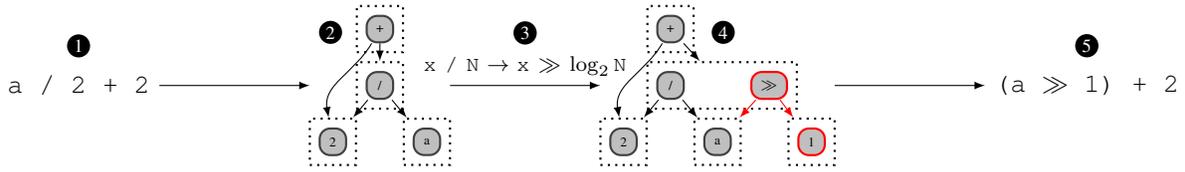
\begin{figure*}
  \centering
  \begin{tikzpicture}[%
    auto,
    block/.style={
      rectangle,
      draw=darkgray,
      thick,
      fill=lightgray,
      minimum width=1em,
      align=center,
      rounded corners,
      minimum height=1em,
      font=\tiny
    },
    line/.style={
      draw,thick,
      -latex',
      shorten >=2pt
    }
  ]

  \node[name=initial-expr, draw=none] {\texttt{a / 2 + 2}};

  \scopenode[name=initial-egraph, draw=none, anchor=west, xshift=-2cm, at=(initial-expr.east)] {
      \draw (0.625,-0.75) node[block] (add) {+};
      \node[thick,draw,dotted,fit=(add)] (addclass) {};
      \draw (0.625,-1.5) node[block] (div) {/};
      \node[thick,draw,dotted,fit=(div)] (divclass) {};
      \draw (1.25,-2.25) node[block] (a) {a};
      \node[thick,draw,dotted,fit=(a)] (aclass) {};
      \draw (0,-2.25) node[block] (two) {2};
      \node[thick,draw,dotted,fit=(two)] (twoclass) {};
      \draw[-latex] (add) -- (divclass);
      \draw[-latex] (add) to[out=240,in=100] (twoclass);
      \draw[-latex] (div) -- (aclass);
      \draw[-latex] (div) -- (twoclass);
  };

  \scopenode[name=updated-egraph, draw=none, anchor=west, xshift=-2cm, at=(initial-egraph.east)] {
      \draw (0.625,-0.75) node[block] (add) {+};
      \node[thick,draw,dotted,fit=(add)] (addclass) {};
      \draw (0.625,-1.5) node[block] (div) {/};
      \draw (1.875,-1.5) node[block,draw=red] (shift) {$\gg$};
      \node[thick,draw,dotted,fit=(div) (shift)] (divclass) {};
      \draw (1.25,-2.25) node[block] (a) {a};
      \node[thick,draw,dotted,fit=(a)] (aclass) {};
      \draw (0,-2.25) node[block] (two) {2};
      \node[thick,draw,dotted,fit=(two)] (twoclass) {};
      \draw (2.5,-2.25) node[block,draw=red] (one) {1};
      \node[thick,draw,dotted,fit=(one)] (oneclass) {};
      \draw[-latex] (add) -- (divclass);
      \draw[-latex] (add) to[out=240,in=100] (twoclass);
      \draw[-latex] (div) -- (aclass);
      \draw[-latex] (div) -- (twoclass);
      \draw[-latex,red] (shift) -- (aclass);
      \draw[-latex,red] (shift) -- (oneclass);
  };

  \node[name=extracted-expr, draw=none, anchor=west, xshift=2cm, at=(updated-egraph.east)] {\texttt{(a $\gg$ 1) + 2}};

  \node[draw=none, anchor=south, at=(initial-expr.north)] {\circled{1}};
  \node[draw=none, yshift=-1em, at=(initial-egraph.north west)] {\circled{2}};
  \node[draw=none, yshift=-1em, xshift=-5em, at=(updated-egraph.north east)] {\circled{4}};
  \node[draw=none, anchor=south, at=(extracted-expr.north)] {\circled{5}};

  \draw[-latex] (initial-expr) -- (initial-egraph);
  \draw[-latex] (initial-egraph) -- node[above, text width=3cm, align=center] {\circled{3} \\ {\footnotesize $\texttt{x / N} \rightarrow \texttt{x $\gg$ $\log_2\texttt{N}$}$}} (updated-egraph);
  \draw[-latex] (updated-egraph) -- (extracted-expr);
\end{tikzpicture}
  \caption{
    Expression \circled{1} is converted to \egraph \circled{2}, which is subsequently saturated.
    In this example, only rule \circled{3} is applied: $\typeconstraints{\forall \texttt{x} \in \mathbb{N}, \texttt{N} \in \mathbb{N}} \texttt{x / N} \rightarrow \texttt{x $\gg$ $\log_2\texttt{N}$}$. %
    From saturated \egraph \circled{4}, expression \circled{5} is selected by an extractor that prefers bitwise shift.
  }
  \label{fig:equality-saturation-workflow}
\end{figure*}

To summarize, this paper makes the following contributions:
\begin{itemize}
    \item Presents a minimalist \ac{IR} and its tiny subset of rewrites suitable for capturing the language semantics;    
    \item Shows how this minimalist \ac{IR} can be used to express idioms found in BLAS and PyTorch;
    \item Demonstrates the effectiveness of using equality saturation with a minimalist \ac{IR} on a set of computational kernels.
\end{itemize}

The rest of this paper is organized as follows:
\Cref{sec:background} introduces equality saturation while \cref{sec:overview} provides an overview of the technique proposed.
\Cref{sec:minIR} introduces the minimalist \ac{IR} and the core rewrite rules.
\Cref{sec:optimizing-ir} shows two use-cases based on \ac{BLAS} and PyTorch.
\Cref{sec:evaluation} evaluates the approach. %
Finally \cref{sec:related} discusses related work and \cref{sec:conclusion} concludes.

\section{Background: Equality Saturation}
\label{sec:background}

Equality saturation is a rule-based rewriting algorithm that explores all variants~\cite{tate2009equality} of an input program.
These variants arise from applying rewrite rules as a fixed-point iteration on a dedicated data structure: the \egraph.
Once the \egraph is built, a performance model extracts a single expression.

\paragraph{\eGraphs}

Equality saturation engines store program variants in a specialized data structure called an \egraph.
Each \egraph consists of a set of \enodes and an equivalence relation that partitions the nodes into \eclasses.
\eNodes have \eclasses as children, allowing each \enode and \eclass to represent a possibly unbounded set of expressions.

Encoding a program as an \egraph is straightforward.
Each expression node becomes an \enode and each unique \enode is placed in its own \eclass.
This is illustrated in \cref{fig:equality-saturation-workflow} for expression \circled{1},  %
and its corresponding \egraph \circled{2}.
Multiple occurrences of \texttt{2} are represented by multiple incoming edges. %

\paragraph{Saturation}

Once an expression has been converted to an \egraph, the graph is saturated.
Saturation repeatedly applies rewrite rules to all eligible nodes in the graph.
Rule application can be performed in any order, but is more efficient when performed in batch~\cite{willsey2021egg}.
A batch consists of all currently possible rewrites.
Applying such a batch is referred to as a \textit{saturation step} or simply \textit{step} in this paper.

In the simplest of cases, a step consists of a single rewrite rule application.
This is the case in \cref{fig:equality-saturation-workflow}, where \egraph \circled{2} is expanded by applying rule \circled{3}: $\texttt{x/N} \rightarrow \texttt{x $\gg$ $\log_2\texttt{N}$}$, where $\gg$ represents the right-shift operator.

Assuming the set of rules only contains this one rule, there are no further possible applications of this rule in \egraph \circled{4}; hence the \egraph is said to have reached a fixpoint.
Such a fixpoint may not always exist, in such cases standard practice is to terminate equality saturation using a timeout~\cite{chandrakana2020synthesizing}.

\paragraph{Extraction}

After reaching either a fixpoint or timeout, \egraph \circled{4} is ready for extraction.
Extraction reduces an \eclass or \enode%
to a single expression.
This single output expression corresponds to a walk through the \egraph, starting from an \enode or \eclass.
Each time the traversal encounters an \eclass, a single \enode is selected from that \eclass.

To guide the walk through the \egraph, one typically looks for the best expression in the \eclass or \enode, based on some measure of quality using a cost function.
To illustrate the use of a cost function, we could define such a function to assign a lower cost to a bitwise shift than to integer division.
This difference in cost gives rise to an extractor that selects expression \circled{5}, \texttt{(a $\gg$ 1) + 2}, from the \egraph.

\section{Overview}
\label{sec:overview}

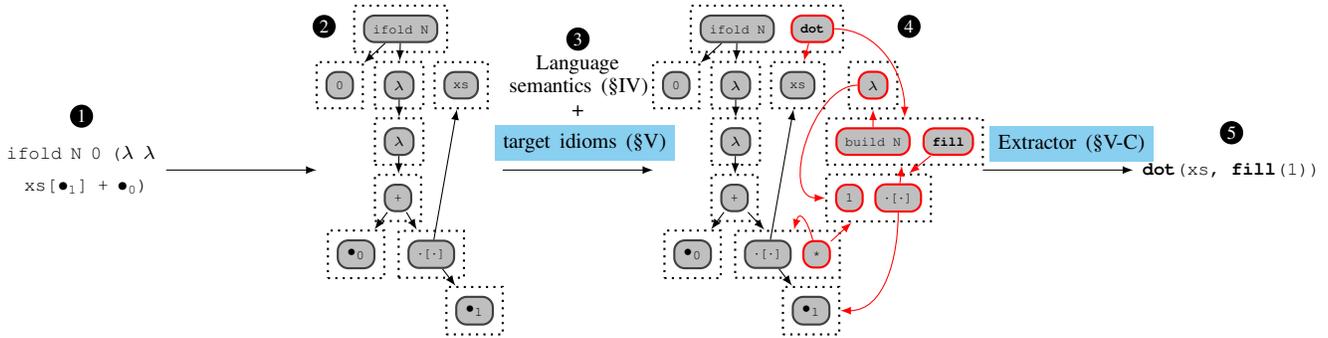
\begin{figure*}
  \centering
  \begin{tikzpicture}[%
    auto,
    block/.style={
      rectangle,
      draw=darkgray,
      thick,
      fill=lightgray,
      minimum width=1em,
      align=center,
      rounded corners,
      minimum height=1em
    },
    line/.style={
      draw,thick,
      -latex',
      shorten >=2pt
    }
  ]

  \node[name=input-expression, text width=2cm] {\scriptsize \texttt{ifold~N~0~(\debruijnlambda{\debruijnlambda{\\\ \ xs[\debruijnindex{1}] + \debruijnindex{0}}})}};

  \scopenode[name=initial-egraph, draw=none, anchor=west, xshift=-2cm, at=(input-expression.east)] {
      \draw (0,0) node[block] (ifold) {\tiny \texttt{ifold N}};
      \node[thick,draw,dotted,fit=(ifold)] (ifoldclass) {};
      \draw (0.275,-0.75) node[block] (lambda1) {\tiny $\lambda$};
      \node[thick,draw,dotted,fit=(lambda1)] (lambda1class) {};
      \draw (-0.5,-0.75) node[block] (zero) {\tiny \texttt{0}};
      \node[thick,draw,dotted,fit=(zero)] (zeroclass) {};
      \draw (0.275,-1.5) node[block] (lambda2) {\tiny $\lambda$};
      \node[thick,draw,dotted,fit=(lambda2)] (lambda2class) {};
      \draw (0.275,-2.25) node[block] (add) {\tiny \texttt{+}};
      \node[thick,draw,dotted,fit=(add)] (addclass) {};
      \draw (-0.35,-3) node[block] (p0) {\tiny \debruijnindex{0}};
      \node[thick,draw,dotted,fit=(p0)] (p0class) {};
      \draw node[block, right of=p0] (index) {\tiny \texttt{$\cdot$[$\cdot$]}};
      \node[thick,draw,dotted,fit=(index)] (indexclass) {};
      \draw node[block, right of=lambda1, xshift=-0.5em] (xs) {\tiny \texttt{xs}};
      \node[thick,draw,dotted,fit=(xs)] (xsclass) {};
      \draw (1.225,-3.75) node[block] (p1) {\tiny \debruijnindex{1}};
      \node[thick,draw,dotted,fit=(p1)] (p1class) {};
      \draw[-latex] (ifold) -- (lambda1class);
      \draw[-latex] (ifold) -- (zeroclass);
      \draw[-latex] (lambda1) -- (lambda2class);
      \draw[-latex] (lambda2) -- (addclass);
      \draw[-latex] (add) -- (indexclass);
      \draw[-latex] (add) -- (p0class);
      \draw[-latex] (index) -- (xsclass);
      \draw[-latex] (index) -- (p1class);
  };

  \scopenode[name=updated-egraph, draw=none, anchor=west, xshift=-2cm, at=(initial-egraph.east)] {
      \draw (0,0) node[block] (ifold) {\tiny \texttt{ifold N}};
      \draw node[block, draw=red, right of=ifold] (dot) {\tiny \apiname{dot}};
      \node[thick,draw,dotted,fit=(ifold) (dot)] (ifoldclass) {};
      \draw (0.275,-0.75) node[block] (lambda1) {\tiny $\lambda$};
      \node[thick,draw,dotted,fit=(lambda1)] (lambda1class) {};
      \draw (-0.5,-0.75) node[block] (zero) {\tiny \texttt{0}};
      \node[thick,draw,dotted,fit=(zero)] (zeroclass) {};
      \draw (0.275,-1.5) node[block] (lambda2) {\tiny $\lambda$};
      \node[thick,draw,dotted,fit=(lambda2)] (lambda2class) {};
      \draw (0.275,-2.25) node[block] (add) {\tiny \texttt{+}};
      \node[thick,draw,dotted,fit=(add)] (addclass) {};
      \draw node[block, draw=red, right of=add, xshift=1.5em] (one) {\tiny \texttt{1}};
      \draw node[block, draw=red, right of=one, xshift=-1em] (index2) {\tiny \texttt{$\cdot$[$\cdot$]}};
      \node[thick,draw,dotted,fit=(one) (index2)] (oneclass) {};
      \draw (-0.35,-3) node[block] (p0) {\tiny \debruijnindex{0}};
      \node[thick,draw,dotted,fit=(p0)] (p0class) {};
      \draw node[block, right of=p0] (index) {\tiny \texttt{$\cdot$[$\cdot$]}};
      \draw node[block, draw=red, xshift=-1em, right of=index] (mul) {\tiny \texttt{*}};
      \node[thick,draw,dotted,fit=(index) (mul)] (indexclass) {};
      \draw node[block, right of=lambda1, xshift=-0.5em] (xs) {\tiny \texttt{xs}};
      \node[thick,draw,dotted,fit=(xs)] (xsclass) {};
      \draw (1.225,-3.75) node[block] (p1) {\tiny \debruijnindex{1}};
      \node[thick,draw,dotted,fit=(p1)] (p1class) {};
      \draw node[block, draw=red, right of=lambda2, xshift=2.35em] (build) {\tiny \texttt{build N}};
      \draw node[block, draw=red, right of=build] (full) {\tiny \apiname{fill}};
      \node[thick,draw,dotted,fit=(build) (full)] (buildclass) {};
      \draw node[block, draw=red, right of=lambda1, xshift=2.35em] (lambda3) {\tiny $\lambda$};
      \node[thick,draw,dotted,fit=(lambda3)] (lambda3class) {};
      \draw[-latex] (ifold) -- (lambda1class);
      \draw[-latex] (ifold) -- (zeroclass);
      \draw[-latex] (lambda1) -- (lambda2class);
      \draw[-latex] (lambda2) -- (addclass);
      \draw[-latex] (add) -- (indexclass);
      \draw[-latex] (add) -- (p0class);
      \draw[-latex] (index) -- (xsclass);
      \draw[-latex] (index) -- (p1class);
      \draw[-latex,red] (mul) edge [loop above, min distance=0.8em] (indexclass);
      \draw[-latex,red] (mul) -- (oneclass);
      \draw[-latex,red] (index2) to[out=270,in=0] (p1class);
      \draw[-latex,red] (index2) -- (buildclass);
      \draw[-latex,red] (build) -- (lambda3class);
      \draw[-latex,red] (lambda3) to[out=180,in=180] (oneclass);
      \draw[-latex,red] (dot) to[out=0,in=90] (buildclass);
      \draw[-latex,red] (dot) -- (xsclass);
      \draw[-latex,red] (full) -- (oneclass);
  };

  \node[name=extracted-expr, draw=none, anchor=west, xshift=2cm, at=(updated-egraph.east)] {\scriptsize \texttt{\apiname{dot}(xs, \apiname{fill}(1))}};

  \node[draw=none, anchor=south, at=(input-expression.north)] {\circled{1}};
  \node[draw=none, xshift=-0.5em, anchor=north west, at=(initial-egraph.north west)] {\circled{2}};
  \node[draw=none, xshift=-2em, anchor=north east, at=(updated-egraph.north east)] {\circled{4}};
  \node[draw=none, anchor=south, at=(extracted-expr.north)] {\circled{5}};

  \draw[-latex] (input-expression) -- (initial-egraph);
  \draw[-latex] (initial-egraph) -- node[above, text width=2.2cm, align=center] {\circled{3} \\ {\footnotesize Language semantics (\S\ref{sec:minIR}) \\ + \\ \colorbox{highlightblue}{target idioms (\S\ref{sec:optimizing-ir})}}} (updated-egraph);
  \draw[-latex] (updated-egraph) -- node[above, text width=1.8cm, align=center] {\footnotesize \colorbox{highlightblue}{Extractor (\S\ref{sec:cost-model})}} (extracted-expr);
\end{tikzpicture}
  \caption{
    Overview of \techniquename, the proposed technique.
    Vector sum expression \circled{1} is converted to \egraph \circled{2}.
    Equality saturation applies a set of rules \circled{3} to \egraph \circled{2}.
    These rules consist of target-independent language semantics and target-specific idioms.
    Rule application yields updated \egraph \circled{4}.
    From \egraph \circled{4}, a target-specific extractor chooses expression \circled{5}.
    The only target-specific components are the target idioms and extractor, both of which are highlighted in blue.
  }
  \label{fig:overview}
\end{figure*}

\Techniquename uses equality saturation to find idioms in functional array programs. %
\Cref{fig:overview} shows how \techniquename augments the basic equality saturation workflow with a target-independent minimalist array \ac{IR} and its associated language semantics rules.
The \ac{IR} is the common representation at every step of the system, from input expression \circled{1} through \egraphs \circled{2} and \circled{4} to final extracted expression \circled{5}.
To allow \techniquename to target libraries, it carves out two target-specific components: an extractor and idiom rewrite rules.
Those rewrite rules are combined with the language semantics rules in \circled{3}.
Together, they allow the equality saturation engine to expose and exploit latent idioms, as illustrated by the appearance of library function calls in extracted expression \circled{5}.

\subsection{Minimalist Array IR and Rewrite Rules}

\ac{IR} design is an important consideration within equality saturation because the choice of \ac{IR} determines the size of \egraphs and the number of rewrite rules.
Those rules in turn affect the system's feasibility, maintainability, and efficiency.

Recent work~\cite{koehler2021sketch} has applied equality saturation to a Lift-like~\cite{steuwer2017lift} functional programming language rooted in the map-reduce paradigm of array processing.
This paradigm offers expressiveness to programmers and abundant parallelizability to compilers.
Despite these attractive features, a marriage of map-reduce and equality saturation faces two main hurdles: the encoding of \lcalculus and the creation of a comprehensive set of rewrite rules for map-reduce--style operators.
The aforementioned recent work addressed the first hurdle by relying on De Bruijn indices instead of named parameters.

The second challenge was addressed by capturing operator identities using a dazzling \num{156} rewrite rules, requiring over \num{1000} lines of Scala code!\footnote{See \url{https://github.com/rise-lang/shine/blob/sges/src/main/scala/rise/eqsat/rules.scala}}
Such a large number of rules is undesirable as it required an enormous effort from the system programmer, increases chances of bugs, and makes it hard to understand whether the rule set is complete.
This very large number of rules is a direct consequence of the high number of operators and all their interactions that need to be considered.

This work retains the innovation of De Bruijn indices while replacing the map-reduce paradigm's operators --- map, reduce, concat, zip, $\ldots$ --- with just three fundamental operators: build, ifold and array indexing.
This combination of three operators was originally introduced by work on \ac{DPS}~\cite{shaikhha2017destination} and is sufficiently powerful to model the map-reduce paradigm's plethora of primitives~\cite{lin2022from}.
As we will see, by paring the number of array processing primitives down to three operators, 
the number of rewrite rules that represents a robust subset of the \ac{IR}'s semantics drops down to just eight!

\subsection{Target-Specific Rules and Extractor}

\Techniquename supplements the \ac{IR} and its core rewrite rules with two target-specific components.
The first is a set of rewrite rules that recognize library idioms. %
The second component is an extractor that selects expressions from \egraphs.

A more in-depth discussion of these components is provided in \cref{sec:optimizing-ir}.
That section describes an implementation of the components for two different libraries: \ac{BLAS} and PyTorch.
Since the target-specific components rely on the target-independent \ac{IR}, we first describe the minimalist \ac{IR} design and its rewrite rules in the next section.

\section{Minimalist Array IR and Rewrite Rules}
\label{sec:minIR}

This section describes \techniquename's minimalist functional array \ac{IR} by first introducing the \ac{IR}'s grammar and operators.
The section then proceeds by deriving rewrite rules from language semantics and concludes with some examples.

\subsection{Syntax and Primitives}

\begin{figure}
    \centering
    \footnotesize
    \begin{align*}
        \texttt{e} ::=\hspace{.5em} &\debruijnlambda{\texttt{e}} &&\textcolor{gray}{\textit{lambda abstraction}} \\
        \mid\hspace{.5em} &\texttt{e e} &&\textcolor{gray}{\textit{lambda application}} \\
        \mid\hspace{.5em} &\debruijnindex{i} &&\textcolor{gray}{\textit{parameter use}} \\
        \mid\hspace{.5em} &\texttt{build N f} &&\textcolor{gray}{\textit{array construction}} \\
        \mid\hspace{.5em} &\texttt{e[e]} &&\textcolor{gray}{\textit{array indexing}} \\
        \mid\hspace{.5em} &\texttt{ifold N e e} &&\textcolor{gray}{\textit{iteration with accumulator}} \\
        \mid\hspace{.5em} &\texttt{tuple e e} &&\textcolor{gray}{\textit{tuple creation}} \\
        \mid\hspace{.5em} &\texttt{fst e} \mid \texttt{snd e} &&\textcolor{gray}{\textit{tuple unpacking}} \\
        \mid\hspace{.5em} &\texttt{\apiname{f}($\overline{\texttt{e}}$)} &&\textcolor{gray}{\textit{named function application}}
    \end{align*}
    \caption{
        The grammar describing the minimalist \ac{IR}.
        $\overline{\texttt{e}}$ indicates zero or more instances of \texttt{e}.
        \texttt{N} is a compile-time integer constant and \apiname{f} is an anonymous function.
        The set of available named functions depends on the problem domain.
        For example, \apiname{gemm} is a named function when targeting \ac{BLAS} but not when targeting PyTorch.
    }
    \label{fig:ir-grammar}
\end{figure}

\Cref{fig:ir-grammar} shows the grammar for the proposed minimalist \ac{IR}.
The \ac{IR} consists of four classes of language primitives: \lcalculus with De Bruijn indices, array operations, tuple operations, and named function calls.

\subsubsection{\texorpdfstring{\lCalculus}{Lambda Calculus} with De Bruijn indices}

De Bruijn indices remove the need for named parameters by identifying a lambda's parameter by the number of lambda definitions between the parameter use and the lambda defining the parameter~\cite{de1972lambda}.
For instance, let \debruijnindex{i} denote a De Bruijn index.
$\debruijnlambda{\debruijnindex{0}}$ is equivalent to \texttt{$\lambda$x$\ldotp$~x} and $\debruijnlambda{\debruijnlambda{\debruijnindex{1}}}$ means \texttt{$\lambda$x$\ldotp$~$\lambda$y$\ldotp$~x}.

A consequence of this standardized naming scheme is that semantically equivalent lambdas become syntactically identical.
This syntactic equivalence is, as discussed in related work~\cite{koehler2021sketch}, beneficial in the context of equality saturation because identical expressions correspond to the same \enode in an \egraph.
In short, De Bruijn indices keep \egraphs small.

\subsubsection{Array operations}

The \ac{IR} supplements \lcalculus with three fundamental array operations: \texttt{build}, array indexing, and \texttt{ifold}.

The \texttt{build} operator takes an array length \texttt{N} and a lambda \texttt{f}.
For each index $\texttt{i} \in \left\{0, 1, \dots, N - 1\right\}$, \texttt{build} computes \texttt{f~i} and packages the resulting values in an array:

{
\footnotesize
\begin{equation*}
    \irbuild{N}{f} = \begin{tabular}{|c|c|c|c|}\hline\texttt{f 0} & \texttt{f 1} & \dots & \texttt{f (N - 1)}\\\hline\end{tabular}
\end{equation*}
}

Array indexing is conventional.
Given an array \texttt{a} and an index \texttt{i}, \texttt{a[i]} produces the \texttt{i}th element of \texttt{a}.

{
\footnotesize
\begin{equation*}
    \texttt{(}\begin{tabular}{|c|c|c|c|}\hline$\texttt{a}_\texttt{0}$ & $\texttt{a}_\texttt{1}$ & $\dots$ & $\texttt{a}_\texttt{N-1}$\\\hline\end{tabular}\texttt{)[i]} = \texttt{a}_\texttt{i}
\end{equation*}
}

The \texttt{ifold} operator's main use lies in array aggregation. %
It takes three arguments: a compile-time length \texttt{N}, an initial accumulator value \texttt{init}, and a folding function \texttt{f} that takes both an index and an accumulator value.
This function is applied iteratively according to the following recursive definition:

{
\footnotesize
\begin{align*}
    &\texttt{ifold 0 init f} = \texttt{init} \\
    &\texttt{ifold (N + 1) init f} = \texttt{f N (ifold N init f)}
\end{align*}
}

\subsubsection{Tuple operations}

Arrays capture sequences of homogeneous data structures whereas tuples in the \ac{IR} encode sequences of heterogeneous data.
The \ac{IR} defines only binary tuples, since \textit{n}-ary tuples can be built by nesting binary tuples.
The two main operations related to tuples are tuple construction with \texttt{tuple}, and tuple extraction with \texttt{fst} and \texttt{snd}.
The semantics of these operations are:

{\footnotesize
\begin{align*}
    &\texttt{tuple a b} = \texttt{(a, b)} \\
    &\texttt{fst (a, b)} = \texttt{a} \\
    &\texttt{snd (a, b)} = \texttt{b}
\end{align*}
}

\subsubsection{Named function calls}

The \ac{IR} operators discussed so far support fundamental data types like functions, arrays, and tuples.
To support operations not covered by the aforementioned operators, the \ac{IR} uses named function calls.

Nullary named functions such as \texttt{\apiname{0}()}, \texttt{\apiname{1}()}, \texttt{\apiname{2}()}, \ldots, can be used to model integer and floating-point constants in the \ac{IR}.
For the sake of simplicity, the function call parentheses are omitted when using constants: \eg \texttt{0}, \texttt{1}, \texttt{2}.

A set of binary functions implement standard scalar arithmetic functions such as \texttt{\apiname{+}(a,~b)}.
To make the notation more natural and readable, infix notation is used for these operators, \eg \texttt{a~+~b}.
The \ac{IR} also supports other scalar functions such as comparison, \eg \texttt{a~>~b}.

Named functions can also be used to capture external library calls in the \ac{IR}.
For instance, the PyTorch \texttt{torch.sum(xs)} function represents the sum of a vector's elements and is equivalent to \texttt{ifold~n~0~($\debruijnlambda{\debruijnlambda{\texttt{xs[\debruijnindex{1}]}~+~\debruijnindex{0}}}$)}, where \texttt{n} is \texttt{xs}'s length.
\Cref{sec:optimizing-ir} will dive deeper into the process of finding such equivalences in programs and how that process is used to target various libraries.

\subsection{Language Semantics as Rewrite Rules}

We now derive the eight rewrite rules that capture the relationships between the core \ac{IR} primitives.
We then discuss how these rules are applied.
We also provide examples to illustrate how the rules can simplify expressions and derive new ones.

\subsubsection{Language semantics}

\begin{listing}
    \centering
    \footnotesize
    \begin{align}
        \tag{\textsc{E-BetaReduce}}
        &\typeconstraints{\forall \texttt{e} \in U, \texttt{y} \in T} \texttt{($\debruijnlambda{\texttt{e}}$) y} = \debruijnsubstitute{\texttt{e}}{\texttt{y}} \label{semantics:breduce} \\
        \tag{\textsc{E-IndexBuild}}
        &\typeconstraints{\forall \texttt{f} \in \mathbb{N} \rightarrow T, \texttt{N} \in \mathbb{N}, \texttt{i} \in \mathbb{N} \cap \left[0, \texttt{N}\right)} \texttt{(\irbuild{f}{N})[i] = f i} \label{semantics:build} \\
        \tag{\textsc{E-FstTuple}}
        &\typeconstraints{\forall \texttt{a} \in T_1, \texttt{b} \in T_2} \texttt{fst (tuple a b) = a} \label{semantics:fst} \\
        \tag{\textsc{E-SndTuple}}
        &\typeconstraints{\forall \texttt{a} \in T_1, \texttt{b} \in T_2} \texttt{snd (tuple a b) = b} \label{semantics:snd} \\
        \tag{\textsc{E-FoldInit}}
        &\texttt{ifold 0 init f} = \texttt{init} \label{semantics:ifold-init} \\
        \tag{\textsc{E-FoldStep}}
        &\texttt{ifold (N + 1) init f} = \texttt{f N (ifold N init f)} \label{semantics:ifold-step}
    \end{align}
    \caption{
        Reduction semantics for the minimalist \ac{IR}.
    }
    \label{fig:array-equalities}
\end{listing}

\begin{listing}
    \centering
    \footnotesize
    \begin{align}
        \tag{\textsc{R-BetaReduce}}
        &\texttt{($\debruijnlambda{\texttt{e}}$) y} \rightarrow \debruijnsubstitute{\texttt{e}}{\texttt{y}} \label{core-rules:beta-reduce} \\ 
        \tag{\textsc{R-IntroLambda}}
        &\texttt{e} \rightarrow \texttt{($\debruijnlambda{\texttt{e}\uparrow}$) y} \label{core-rules:intro-lambda}\\
        \tag{\textsc{R-ElimIndexBuild}}
        &\texttt{(\irbuild{f}{N})[i]} \rightarrow \texttt{f i} \label{core-rules:elim-index-build} \\
        \tag{\textsc{R-IntroIndexBuild}}
        &\texttt{f i} \rightarrow \texttt{(\irbuild{f}{N})[i]} \label{core-rules:intro-index-build} \\
        \tag{\textsc{R-ElimFstTuple}}
        &\texttt{fst (tuple a b)} \rightarrow \texttt{a} \label{core-rules:elim-fst-tuple} \\
        \tag{\textsc{R-IntroFstTuple}}
        &\texttt{a} \rightarrow \texttt{fst (tuple a b)} \label{core-rules:intro-fst-tuple} \\
        \tag{\textsc{R-ElimSndTuple}}
        &\texttt{snd (tuple a b)} \rightarrow \texttt{b} \label{core-rules:elim-snd-tuple} \\
        \tag{\textsc{R-IntroSndTuple}}
        &\texttt{b} \rightarrow \texttt{snd (tuple a b)} \label{core-rules:intro-snd-tuple} 
    \end{align}
    \caption{
        Eight rewrite rules that capture the relationships between \texttt{build}, array access, tuple construction, and tuple deconstruction, \labstraction and \breduction.
    }
    \label{fig:rewrite-rules}
\end{listing}

The rewrite rules are obtained from the \ac{IR}'s reduction semantics.
Those semantics are themselves obtained primarily by observing that the \ac{IR} semantics equations from the previous subsection can be conveniently substituted into each other.
Augmenting that substitution with an identity for \breduction results in the set of identities in \cref{fig:array-equalities}.

The first identity corresponds to \breduction where $\debruijnsubstitute{\texttt{e}}{\texttt{y}}$ represents the substitution operator for De Bruijn indices.
This operator transforms an expression by replacing all references to free variable $\debruijnindex{0}$ in \texttt{e} with \texttt{y} and by then lowering the indices of all other free variables in the resulting expression~\cite{de1972lambda}.
For instance, $\debruijnsubstitute{\debruijnindex{0}}{\texttt{y}} = \texttt{y}$ and $\debruijnsubstitute{\debruijnindex{1}}{\texttt{y}} = \debruijnindex{0}$.

\subsubsection{Rewrite rules}

The identities shown above translate readily to the set of rewrite rules in \cref{fig:rewrite-rules}.
Except for \texttt{ifold}, there is one pair of rewrite for each identity.
Although the semantics of \texttt{ifold} could also be expressed as rewrites, the evaluation section will show that for the examples considered, it is not necessary to rewrite any \texttt{ifold}.

All rewrites should be self-explanatory, with the exception of the second one.
The $\uparrow$ operator is called the \textit{shift operator.}
This operator increments the indices of all free variables in \texttt{e} to make room for the additional parameter introduced by the lambda.
For instance, if $\texttt{e} = \debruijnindex{0}$ then $\texttt{($\debruijnlambda{\texttt{e}\uparrow}$) y} = \texttt{($\debruijnlambda{\debruijnindex{1}}$) y}$.

That close relationship with the language semantics guarantees that the rewrite rules capture a robust subset of the \ac{IR}'s semantics.
This subset will prove of interest in \cref{sec:optimizing-ir} and the experiments in \cref{sec:evaluation} will further show that these eight core rules allow an equality saturation engine to effectively reason about and restructure array programs to expose latent idiom occurrences.
These occurrences will then be rewritten as calls to highly optimized library functions.

\subsubsection{Substitution and shift operators}

An interesting property of the rewrite rules is that \ref{core-rules:beta-reduce} and \ref{core-rules:intro-lambda} include operators that are part of the rule application process itself: the substitution operator $\debruijnsubstituteoperator$ and the shift operator $\uparrow$.
These operators manipulate expressions rather than values, making them challenging to express in an equality saturation setting.

This challenge stems from the fact that pattern matching on \egraphs maps the unbound expressions from the rules' left-hand sides to \eclasses.
Each such \eclass captures a potentially unbounded set of expressions whereas substitution and shifting are defined for single expressions.

The literature covers two approaches to address this mismatch.
The first approach lifts the substitution and shift operators into the \egraph and manipulates them with special rewrite rules, but these rules necessitate additional saturation steps and generates wasteful intermediate nodes~\cite{willsey2021egg}.
The second approach applies the operators to individual expressions extracted from each \eclass, which requires only one saturation step and generates no wasteful intermediate nodes~\cite{koehler2021sketch}.
This work makes use of the second technique.

\subsubsection{Free variables in patterns}

Another challenge arises from rules that inflate expressions.
Consider \ref{core-rules:intro-fst-tuple}:

{\footnotesize
\begin{equation*}
    \texttt{a} \rightarrow \texttt{fst (tuple a b)}.
\end{equation*}
}
Standard rule application dictates that whenever the rule matches an \eclass \texttt{a}, an expression \texttt{fst~(tuple~a~b)} is constructed, added to the \egraph, and unified with \texttt{a}.
Most of these steps are unproblematic, but constructing expression \texttt{fst~(tuple~a~b)} is non-obvious because \texttt{b} is an unbound variable.
Intuitively, this means the rule holds for all \texttt{b}.

This work implements that intuition by searching the \egraph for all \eclasses that could serve as \texttt{b}.
In this case, that is every \eclass in the graph.
Expression \texttt{fst~(tuple~a~b)} is then constructed for each suitable \texttt{b}, added to the \egraph, and unified with \texttt{a}.
This approach is applied for every rule where an unbound variable appear on the right-hand side: \ref{core-rules:intro-fst-tuple}, \ref{core-rules:intro-snd-tuple}, \ref{core-rules:intro-index-build} and \ref{core-rules:intro-lambda}.

\subsection{Examples}

To illustrate how the minimalist \ac{IR} and its core rewrite rules work in practice, we consider two examples: map fusion and constant array construction.

\subsubsection{Map fusion}

A standard identity under the map-reduce paradigm is that a pair of map calls can be fused:

{
\footnotesize
\begin{equation*}
    \texttt{map f (map g xs) = map (g $\circ$ f) xs}.
\end{equation*}
}

This assertion is an axiom if \texttt{map} is an irreducible operator.
However, if \texttt{map} is expressed in terms of \texttt{build}, map fusion and fission follows readily from \ref{semantics:build} and \ref{semantics:breduce} as stated in \cref{fig:array-equalities}:

{
\footnotesize
\begin{align*}
    &\texttt{\irbuildf{n}{f (\irbuildf{n}{g xs[\debruijnindex{0}]})[\debruijnindex{0}]}} \\
    &=\texttt{\irbuildf{n}{f ((\debruijnlambda{g xs[\debruijnindex{0}]}) \debruijnindex{0})}} \\
    &= \texttt{\irbuildf{n}{f (g xs[\debruijnindex{0}])}}.
\end{align*}
}

Left to right, the identity also follows from rewrite rules \ref{core-rules:elim-index-build} and \ref{core-rules:beta-reduce}, meaning an equality saturation engine equipped with those rules will find that maps can be fused.
Map fission, the right to left reading of the identity, would require an additional rule due to the specific \labstraction rule in use.
We choose not to include such a rule because it will not be of interest to the evaluation in \cref{sec:evaluation}.

\subsubsection{Constant array construction}

Suppose that we would like to optimize the following expression which adds $42$ to each element of the array \texttt{xs}:

{
\footnotesize
\begin{equation*}
    \texttt{\irbuildf{n}{xs[\debruijnindex{0}] + 42}}.
\end{equation*}
}

Let us assume that we have a library available with very fast implementations of \apiname{addvec}, which computes the elementwise addition of two vectors, and \apiname{constvec}, which creates a vector of constants.
We can express these functions as rewrites using the \techniquename \ac{IR}:

{\footnotesize
\begin{align*}
    & \texttt{\irbuildf{n}{a[\debruijnindex{0}] + b[\debruijnindex{0}])}}  \rightarrow \apiname{addvec(a, b)}, \\
    &\texttt{\irbuildf{n}{c}} \rightarrow \apiname{constvec(c)}.
\end{align*}
}

Unfortunately, neither of these patterns appear in the expression to optimize.
However, the rewrite rules seen earlier allow for a scalar to be transformed to an indexed array of such scalars:

{
\footnotesize
\begin{equation*}
    \texttt{0} = \texttt{($\debruijnlambda{\texttt{0}}$)}~\texttt{i} = \texttt{\irbuildf{n}{\texttt{0}}[i]}.
\end{equation*}
}

This identity allows us to infer that

{
\footnotesize
\begin{align*}
    &\irbuildf{n}{\texttt{xs[\debruijnindex{0}] + 42}} \\
    &= \irbuildf{n}{\texttt{xs[\debruijnindex{0}] + (\irbuildf{n}{42})[\debruijnindex{0}])}} \\
    &= \texttt{\apiname{addvec}(xs, \irbuildf{n}{42})} \\
    &= \texttt{\apiname{addvec}(xs, \apiname{constvec}(42))}.
\end{align*}
}

To summarize, this process is fully automated by simply encoding once and for all the library functions idioms as rewrite rules.
Coupled with the eight core rewrite rules presented earlier, the equality saturation engine can simply do its job and find the \emph{latent} idioms --- they are not directly observable --- in the original expression.
Once the idioms have been identified, the original application can be readily accelerated using one or more calls to the high-performance library.

\section{Use Cases: BLAS and PyTorch}
\label{sec:optimizing-ir}

The constant array construction example from the previous section illustrates that \techniquename's \ac{IR} is suitable for robust pattern matching and rewriting, even when input programs do not exactly match library idioms.
This section generalizes the approach from that example and implements it more rigorously for two libraries: \ac{BLAS} and PyTorch.
Those implementations follow the blueprint from \cref{fig:overview} in \cref{sec:overview}; hence, they consist of two components: target idioms and an extractor.

Since \ac{BLAS} and PyTorch both operate on floating-point numbers, this section separates the idiom rules into shared scalar arithmetic rules and library-specific idiom rules.
The section then presents the extractor and its cost model, which also facilitates sharing between the \ac{BLAS} and PyTorch targets.

\subsection{Scalar Arithmetic Rules}

\begin{listing}
    \centering
    \footnotesize
    \begin{align}
        \tag{\textsc{E-AddZero}}
        &\typeconstraints{\forall \texttt{x} \in \mathbb{R}} \texttt{x + 0} = \texttt{x}
        \label{scalar-rules:add-zero} \\
        \tag{\textsc{E-MulOneL}}
        &\typeconstraints{\forall \texttt{x} \in \mathbb{R}} \texttt{1 * x} = \texttt{x}
        \label{scalar-rules:mul-one-left} \\
        \tag{\textsc{E-MulOneR}}
        &\typeconstraints{\forall \texttt{x} \in \mathbb{R}} \texttt{x * 1} = \texttt{x}
        \label{scalar-rules:mul-one-right} \\
        \tag{\textsc{E-CommuteMul}}
        &\typeconstraints{\forall \texttt{x} \in \mathbb{R}, \texttt{y} \in \mathbb{R}} \texttt{x * y} = \texttt{y * x}
        \label{scalar-rules:commute-mul}
    \end{align}
    \caption{
        Scalar arithmetic identities.
        Each identity corresponds to two rewrite rules: a left-to-right rule and a right-to-left rule.
        \texttt{x} and \texttt{y} are numbers.
    }
    \label{lst:scalar-rules}
\end{listing}

\Cref{lst:scalar-rules} provides a small list of scalar rewrite rules.
When combined with the core rules, the scalar rules allow an equality saturation engine to reason about tensors.
That reasoning is designed to enable idiom detection even if the idioms seem hidden at first.

For example, consider the latent dot product in vector sum

{
\footnotesize
\begin{equation*}
    \texttt{ifold n 0 (\debruijnlambda{\debruijnlambda{xs[\debruijnindex{1}] + \debruijnindex{0}}})}.
\end{equation*}
}

We can expose that dot product idiom by first applying \ref{scalar-rules:mul-one-right} to \texttt{xs[\debruijnindex{1}]}, yielding \texttt{xs[\debruijnindex{1}] * 1}.
We then use \ref{core-rules:intro-lambda} and \ref{core-rules:intro-index-build} on that constant \texttt{1} to obtain a final expression of

{
\footnotesize
\begin{align*}
    &\texttt{ifold n 0 (\debruijnlambda{\debruijnlambda{xs[\debruijnindex{1}] * (\irbuildf{n}{1})[\debruijnindex{1}] + \debruijnindex{0}}})} 
\end{align*}
}

which is equivalent to a dot product with a vector of ones:  \texttt{\apiname{dot}(xs, \irbuildf{n}{1})}.

\subsection{Idiom Rules}

The dot product idiom recognition example illustrates how the core and scalar rewrite rules work together to expose latent idioms.
Once such idioms have been exposed, they still need to be recognized as such.
To that end, we now introduce idiom-recognizing rewrite rules for \ac{BLAS} and PyTorch.

\subsubsection{BLAS}

\begin{listing}
\centering\footnotesize
\begin{align*}
    &\texttt{\apiname{axpy}(alpha, A, B)} \\
    \tag{\textsc{I-Axpy}}
    &= \texttt{\irbuildf{N}{alpha$\uparrow$ * A$\uparrow$[\debruijnindex{0}] + B$\uparrow$[\debruijnindex{0}]}}
    \label{idiom:blas-axpy} \\
    &\texttt{\apiname{dot}(A, B)} \\
    \tag{\textsc{I-Dot}}
    &= \texttt{ifold N 0 (\debruijnlambda{\debruijnlambda{A$\uparrow\uparrow$[\debruijnindex{1}] * B$\uparrow\uparrow$[\debruijnindex{1}] + \debruijnindex{0}}})}
    \label{idiom:blas-dot} \\
    &\texttt{\gemv{F}(alpha, A, B, beta, C)} \\
    \tag{\textsc{I-Gemv}}
    &= \texttt{\irbuildf{N}{alpha$\uparrow$ * \apiname{dot}(A$\uparrow$[\debruijnindex{0}], B$\uparrow$) + beta$\uparrow$ * C$\uparrow$[\debruijnindex{0}]}}
    \label{idiom:blas-gemv} \\
    &\texttt{\gemm{F}{T}(alpha, A, B, beta, C)} \\
    \tag{\textsc{I-Gemm}}
    &= \texttt{\irbuildf{N}{\gemv{N}(alpha$\uparrow$, B$\uparrow$, A$\uparrow$[\debruijnindex{0}], beta$\uparrow$, C$\uparrow$[\debruijnindex{0}])}}
    \label{idiom:blas-gemm} \\
    &\texttt{\apiname{transpose}(A)} \\
    \tag{\textsc{I-Transpose}}
    &= \texttt{\irbuildf{N}{\irbuildf{M}{A$\uparrow\uparrow$[\debruijnindex{0}][\debruijnindex{1}]}}}
    \label{idiom:blas-transpose} \\
    &\texttt{\gemv{X}(alpha, \apiname{transpose}(A), B, beta, c)} \\
    \tag{\textsc{I-TransposeInGemv}}
    &=\texttt{\gemv{$\neg$X}(alpha, A, B, beta, c)}
    \label{idiom:blas-fold-transpose-gemv} \\
    &\texttt{\gemm{X}{Y}(alpha, \apiname{transpose}(A), B, beta, C)} \\
    \tag{\textsc{I-TransposeAInGemm}}
    &=\texttt{\gemm{$\neg$X}{Y}(alpha, A, B, beta, C)}
    \label{idiom:blas-fold-transpose-a-into-gemm} \\
    &\texttt{\gemm{X}{Y}(alpha, A, \apiname{transpose}(B), beta, c)} \\
    \tag{\textsc{I-TransposeBInGemm}}
    &=\texttt{\gemm{X}{$\neg$Y}(alpha, A, B, beta, c)}
    \label{idiom:blas-fold-transpose-b-into-gemm} \\
    &\texttt{\apiname{dot}(\irbuildf{N}{alpha * A[\debruijnindex{0}]}, B)} \\
    \tag{\textsc{I-HoistMulFromDot}}
    &=\texttt{alpha * \apiname{dot}(A, B)}
    \label{idiom:blas-hoist-mul-from-dot} \\
    &\texttt{\apiname{memset}(0)} \\
    \tag{\textsc{I-MemsetZero}}
    &= \texttt{\irbuildf{N}{0}}
    \label{idiom:blas-memset-zero}
\end{align*}
\caption{
    BLAS idioms considered in this work.
    \texttt{F} in \gemv{F} is short for \textit{false} and indicates that matrix \texttt{A} is not transposed.
    \texttt{F,T} in \gemm{F}{T} indicate that \texttt{A} is not transposed and \texttt{B} is transposed.
}
\label{lst:blas-idioms}
\end{listing}

\Cref{lst:blas-idioms} contains equivalences that define idioms corresponding to five \ac{BLAS} functions.
These functions are:
\begin{enumerate}
    \item \apiname{dot}: computes the dot product of two vectors;
    \item \apiname{axpy}: computes $\alpha A + B$, where $\alpha$ is a scalar and $A$, $B$ are vectors;
    \item \gemv{X}: computes $\alpha A B + \beta C$, where $\alpha$, $\beta$ are scalars, $B$, $C$ are vectors, and $A$ is a matrix;
    \item \gemm{X}{Y}: computes $\alpha A B + \beta C$, where $\alpha$, $\beta$ are scalars and $A$, $B$, $C$ are matrices; and
    \item \apiname{transpose}: transposes a matrix. %
\end{enumerate}

\begin{listing}
\centering\footnotesize
\begin{align*}
    &\texttt{\apiname{dot}(A, B)} \\
    \tag{\textsc{I-Dot}}
    &= \texttt{ifold N 0 (\debruijnlambda{\debruijnlambda{A$\uparrow\uparrow$[\debruijnindex{1}] * B$\uparrow\uparrow$[\debruijnindex{1}] + \debruijnindex{0}}})}
    \label{idiom:torch-dot} \\
    &\texttt{\apiname{sum}(A)} \\
    \tag{\textsc{I-VecSum}}
    &= \texttt{ifold N 0 (\debruijnlambda{\debruijnlambda{A$\uparrow\uparrow$[\debruijnindex{1}] + \debruijnindex{0}}})}
    \label{idiom:torch-reduce-sum} \\
    &\texttt{\apiname{mv}(A, B)} \\
    \tag{\textsc{I-MatVec}}
    &= \texttt{\irbuildf{N}{\apiname{dot}(A$\uparrow$[\debruijnindex{1}], B$\uparrow$)}}
    \label{idiom:torch-matvec} \\
    &\texttt{\apiname{mm}(A, B)} \\
    \tag{\textsc{I-MatMat}}
    &= \texttt{\irbuildf{N}{\apiname{mv}(B$\uparrow$, A$\uparrow$[\debruijnindex{1}])}}
    \label{idiom:torch-matmul} \\
    &\texttt{\apiname{transpose}(A)} \\
    \tag{\textsc{I-Transpose}}
    &= \texttt{\irbuildf{N}{\irbuildf{M}{A$\uparrow\uparrow$[\debruijnindex{0}][\debruijnindex{1}]}}}
    \label{idiom:torch-transpose} \\
    &\texttt{\apiname{transpose}(\apiname{transpose}(A))} \\
    \tag{\textsc{I-TransposeTwice}}
    &= \texttt{A}
    \label{idiom:torch-transpose-twice} \\
    &\texttt{\apiname{add}(A, B)} \\
    \tag{\textsc{I-AddVec}}
    &= \texttt{\irbuildf{N}{A$\uparrow$[\debruijnindex{0}] + B$\uparrow$[\debruijnindex{0}]}}
    \label{idiom:torch-add-vec} \\
    &\texttt{\apiname{add}(A, B)} \\
    \tag{\textsc{I-LiftAdd}}
    &= \texttt{\irbuildf{N}{\apiname{add}(A$\uparrow$[\debruijnindex{0}], B$\uparrow$[\debruijnindex{0}])}}
    \label{idiom:torch-lift-add} \\
    &\texttt{\apiname{mul}(alpha, A)} \\
    \tag{\textsc{I-MulScalarAndVec}}
    &= \texttt{\irbuildf{N}{alpha * A$\uparrow$[\debruijnindex{0}]}}
    \label{idiom:torch-mul-scalar-and-vec} \\
    &\texttt{\apiname{mul}(alpha, A)} \\
    \tag{\textsc{I-LiftMul}}
    &= \texttt{\irbuildf{N}{\apiname{mul}(alpha, A$\uparrow$[\debruijnindex{0}])}}
    \label{idiom:torch-lift-mul} \\
    &\texttt{\apiname{full}(c)} \\
    \tag{\textsc{I-FullVec}}
    &= \texttt{\irbuildf{N}{c$\uparrow$}}
    \label{idiom:torch-fill-vec} %
\end{align*}
\caption{
    PyTorch idioms considered in this work.
    \ref{idiom:torch-transpose-twice} captures a property of the \apiname{transpose} function; all other rules recognize idioms. 
}
\label{lst:pytorch-idioms}
\end{listing}

The idioms in \cref{lst:blas-idioms} encompass both the transposed and non-transposed variants of these \ac{BLAS} functions, resulting in a larger number of equivalences.
For instance, in the case of \gemv{X}, \ref{idiom:blas-gemv} defines \gemv{F}, and \ref{idiom:blas-fold-transpose-gemv} connects \gemv{F} to its transposed counterpart, denoted as \gemv{T}.
The inclusion of these transposed variations allows for a more comprehensive definition of \ac{BLAS} functions.
Additionally, \cref{lst:blas-idioms} also presents an idiom for the C standard library \apiname{memset} function, which can be used to quickly create an all-zeros vector.
The shift operator $\left(\uparrow\right)$ applications in \cref{lst:blas-idioms} increment De Bruijn indices to avoid overlap with new parameters introduced by \labstraction.

\subsubsection{PyTorch}

We implement a similar set of idioms for PyTorch, described in \cref{lst:pytorch-idioms}.
The PyTorch functions corresponding to those idioms take fewer parameters than their \ac{BLAS} counterparts, but are often polymorphic.
For instance, \texttt{A} in \texttt{\apiname{mul}(alpha, A)} could be a scalar, vector, matrix or higher-order tensor.
An array of multiplications \texttt{[mul(a, A$_1$), mul(a, A$_2$), \dots]} can hence be rewritten as a single call \texttt{mul(a, [A$_1$, A$_2$, \dots])}.
\ref{idiom:torch-lift-mul} captures this polymorphic property by defining a higher-dimensional \apiname{mul} as a vector of lower-dimensional \apiname{mul} calls.
\ref{idiom:torch-lift-add} accomplishes the same for the \apiname{add} function. 

\subsection{Cost Model}
\label{sec:cost-model}

The extraction step of equality saturation, initially designed for a pseudo-Boolean solver, can be implemented in various ways~\cite{tate2009equality}.
The most popular implementation employs a local cost model, which quickly and simply calculates the cost of each \enode within an \eclass based on its arguments' cost~\cite{willsey2021egg}.
An \eclass's cost is determined by its cheapest \enode, and the extraction process involves selecting the cheapest \enode recursively from an \eclass.

This paper opts for the cost model-based approach for simplicity's sake --- the extractor for the \ac{BLAS} and PyTorch use cases is not the focal point of this work.
The cost models for the two use cases consist of a common base and a library function %
cost model.
The common base cost in \cref{def:cost-function-common} describes the cost of the core \ac{IR} operators and library-independent named functions.
\Cref{def:cost-function-blas} and \cref{def:cost-function-pytorch}  capture the \ac{BLAS} and PyTorch library function cost model respectively.

\begin{listing}
\footnotesize
\begin{align*}
    &\costof{\irbuild{N}{f}} &&= \texttt{N} \cdot \left(\costof{\texttt{f}} + 1\right) + 1 \\
    &\costof{\texttt{A[i]}} &&= \costof{\texttt{A}} + \costof{\texttt{i}} + 1 \\
    &\costof{\texttt{ifold N init f}} &&= \costof{\texttt{init}} + \texttt{N} \cdot \costof{\texttt{f}} + 1 \\
    &\costof{\texttt{tuple a b}} &&= \costof{\texttt{a}} + \costof{\texttt{b}} + 1 \\
    &\costof{\texttt{fst t}} &&= \costof{\texttt{t}} + 1 \\
    &\costof{\texttt{snd t}} &&= \costof{\texttt{t}} + 1 \\
    &\costof{\debruijnlambda{\texttt{e}}} &&= \costof{\texttt{e}} + 1 \\
    &\costof{\texttt{f~e}} &&= \costof{\texttt{f}} + \costof{\texttt{e}} + 1 \\
    &\costof{\debruijnindex{k}} &&= 1 \left(\forall \texttt{k} \in \mathbb{N}\right) \\
    &\costof{\texttt{a + b}} &&= \costof{\texttt{a}} + \costof{\texttt{b}} + 1 \\
    &\costof{\texttt{a * b}} &&= \costof{\texttt{a}} + \costof{\texttt{b}} + 1 \\
    &\costof{c} &&= 1 \left(\forall \texttt{c} \in \mathbb{R}\right)
\end{align*}
\caption{
    Definition of the base cost function.
}
\label{def:cost-function-common}
\end{listing}

\begin{listing}
\footnotesize
\begin{align*}
    &\costof{\texttt{\apiname{memset}(c)}} &&= \costof{\texttt{c}} + .8 N + 1 \\
    &\costof{\texttt{\apiname{dot}(A,B)}} &&= \costof{\texttt{A}} + \costof{\texttt{B}} + .8 N \\
    &\costof{\texttt{\apiname{axpy}(a,A,B)}} &&= \costof{\texttt{a}} + \dots + \costof{\texttt{B}} + .8 N \\
    &\costof{\texttt{\apiname{gemv}(a,A,B,b,C)}} &&= \costof{\texttt{a}} + \dots + \costof{\texttt{C}} + .7 N M \\
    &\costof{\texttt{\apiname{gemm}(a,A,B,b,C)}} &&= \costof{\texttt{a}} + \dots + \costof{\texttt{C}} + .6 N M K \\
    &\costof{\texttt{\apiname{transpose}(A)}} &&= \costof{\texttt{A}} + .9 N M \\
\end{align*}
\caption{
    \ac{BLAS}-specific additions to $\costfunc$.
    Calls to external functions are discounted to make them more attractive.
    Discounting factors are chosen semi-arbitrarily.
    $N$, $M$, and $K$ are array dimensions.
}
\label{def:cost-function-blas}
\end{listing}

\begin{listing}
\footnotesize
\begin{align*}
    &\costof{\texttt{\apiname{full}(c)}} &&= \costof{\texttt{c}} + .8 N + 1 \\
    &\costof{\texttt{\apiname{add}(A, B)}} &&= \costof{\texttt{A}} + \costof{\texttt{B}} + .4 N + .4 M \\
    &\costof{\texttt{\apiname{mul}(A, B)}} &&= \costof{\texttt{A}} + \costof{\texttt{B}} + .4 N + .4 M \\
    &\costof{\texttt{\apiname{sum}(A, B)}} &&= \costof{\texttt{A}} + \costof{\texttt{B}} + .8 N \\
    &\costof{\texttt{\apiname{dot}(A, B)}} &&= \costof{\texttt{A}} + \costof{\texttt{B}} + .8 N \\
    &\costof{\texttt{\apiname{mv}(A, B)}} &&= \costof{\texttt{A}} + \costof{\texttt{B}} + .7 N M \\
    &\costof{\texttt{\apiname{mm}(A, B)}} &&= \costof{\texttt{A}} + \costof{\texttt{B}} + .6 N M K \\
    &\costof{\texttt{\apiname{transpose}(A)}} &&= \costof{\texttt{A}} + .9 N M %
\end{align*}
\caption{
    PyTorch-specific additions to $\costfunc$.
    $N$ and $M$ are array dimensions.
    For polymorphic arrays, $N$ and $M$ represent the product of the arrays' dimensions.
}
\label{def:cost-function-pytorch}
\end{listing}

\section{Evaluation}
\label{sec:evaluation}

The previous section has examined qualitatively how \techniquename can target different libraries such as \ac{BLAS} and PyTorch.
This section uses quantitative experiments to measure how well \techniquename identifies idioms and speeds up programs.

To perform these experiments, we implement the \ac{IR} as a Scala \ac{DSL}.
We encode in that \ac{DSL} a subset of the PolyBench/C 4.2.1-beta benchmark suite~\cite{pouchet2016polybench} and add custom kernels to evaluate specific tasks.
The custom and PolyBench kernels are described in \cref{tab:benchmark-descriptions}.

Kernels are expressed by composing \texttt{build}-\texttt{ifold} implementations of the respective mathematical operators as in prior work~\cite{shaikhha2017destination}.
For instance, \benchmarkname{gemv} becomes:

{
\footnotesize
\begin{align*}
    &\texttt{gemv($\alpha$, A, B, $\beta$, C)} \\
    &= \texttt{vadd(vscale($\alpha$, matvec(A, B)), vscale($\beta$, C))}.
\end{align*}
}

\texttt{vadd}, \texttt{vscale} and \texttt{matvec} expand as below.

{
\footnotesize
\begin{align*}
    &\texttt{vadd(A, B) = \irbuildf{N}{A$\uparrow$[\debruijnindex{0}] + B$\uparrow$[\debruijnindex{0}]}} \\
    &\texttt{vscale($\alpha$, A) = \irbuildf{N}{$\alpha$$\uparrow$ * A$\uparrow$[\debruijnindex{0}]}} \\
    &\texttt{matvec(A, B) = \irbuildf{N}{dot(A$\uparrow$[\debruijnindex{0}], B$\uparrow$)}} \\
    &\texttt{dot(A, B) = \texttt{ifold N 0 (\debruijnlambda{\debruijnlambda{A$\uparrow\uparrow$[\debruijnindex{1}] * B$\uparrow\uparrow$[\debruijnindex{1}] + \debruijnindex{0}}})}}
\end{align*}
}

The exceptions to this scheme are \benchmarkname{doitgen} and \benchmarkname{gemver}.
We translate their C loops directly to \texttt{build} and \texttt{ifold}.

\begin{table}
    \centering
    \footnotesize
    \caption{
        Overview of kernels examined in this work.
        PolyBench kernel descriptions adapted from benchmark suite~\cite{pouchet2016polybench}.
    }
    \label{tab:benchmark-descriptions}
    \begin{tabular}{l l l}%
        \hline
        \textbf{Kernel} & \textbf{Suite} & \textbf{Description} \\ %
        \hline
        2mm & PolyBench & Two generalized matrix multiplications \\
        atax & PolyBench & Matrix transpose and vector multiplication \\
        doitgen & PolyBench & Multiresolution analysis kernel (MADNESS) \\
        gemm & PolyBench & Generalized matrix product \\
        gemver & PolyBench & Vector multiplication and matrix addition \\
        gesummv & PolyBench & Scalar, vector and matrix multiplication \\
        jacobi1d & PolyBench & 1D Jacobi stencil computation \\
        mvt & PolyBench & Matrix-vector product and transpose \\
        1mm & Custom & One matrix multiplication \\
        axpy & Custom & Vector scaling and addition \\
        blur1d & Custom & 1D stencil \\
        gemv & Custom & Generalized matrix-vector product \\
        memset & Custom & Zero vector creation \\
        slim-2mm & Custom & Two matrix multiplications \\
        stencil2d & Custom & 2D stencil \\
        vsum & Custom & Vector reduction with sum \\
    \end{tabular}
\end{table}

\textbf{Equality Saturation engine: }
All kernels are transformed by a Scala equality saturation engine %
inspired by the efficient \texttt{egg} implementation~\cite{willsey2021egg}.
The engine performs \egraph construction, saturation and expression extraction.
Saturation proceeds based on one of three sets of rewrite rules, which we term \textit{targets:}
\begin{enumerate}
    \item \textbf{Pure C:} Core and scalar rules only;
    \item \textbf{\ac{BLAS} idioms:} Core, scalar and \ac{BLAS} rules;
    \item \textbf{PyTorch idioms:} Core, scalar and PyTorch rules.
\end{enumerate}
The engine runs equality saturation for five minutes per kernel per benchmark.
After each saturation step, the cost model from \cref{sec:cost-model} selects the optimal expression for that step.

\textbf{Code Generation: }
For \ac{BLAS} and pure C, the selected expressions are compiled to C using an approach similar to prior work~\cite{lin2022from} on C compilation from a functional \ac{IR}.
There are no run-time Python results presented since the compiler used does not currently have a Python back-end.
As such, results for PyTorch are purely qualitative.

We assess \techniquename through experiments on library calls, their evolution over time, code profiling, and run time comparison.

\subsection{Experimental Setup}

We run all experiments on a server with two 18-core Intel Xeon Gold 6254 CPUs and \SI{1}{\tebi\byte} of RAM.
The server uses the following software: CentOS Linux 7, Scala 2.12.7, GCC 11.2.1, and OpenBLAS 0.3.3. %
Through OpenBLAS, compiled kernels take advantage of the server's many cores.
Kernels not compiled for OpenBLAS %
are single-threaded.
The \techniquename implementation that generates kernels is also single-threaded.

\subsection{Idioms Recognized}
\label{sec:evaluation:idioms-recognized}

This section qualitatively assesses how well \techniquename finds \ac{BLAS} and PyTorch idioms.
To perform this assessment, we report %
the library calls found in extracted expressions.
These data are reported for all kernels by \cref{tab:benchmark-overview-blas} and \cref{tab:benchmark-overview-pytorch}.

The tables show that \techniquename finds idioms in each kernel.
The idioms found depend on the kernel and target.
For instance, \techniquename finds that the \benchmarkname{gemv} kernel is best described as a \apiname{gemv} call when targeting \ac{BLAS}.
Indeed, the solution found is simply

{
\footnotesize
\begin{equation*}
\texttt{\gemv{F}($\alpha$, A, B, $\beta$, C)}.
\end{equation*}
}

When targeting PyTorch, \techniquename implements the same kernel as a combination of more granular \texttt{add}, \texttt{mul} and \texttt{mv} calls.

{
\footnotesize
\begin{equation*}
\texttt{\apiname{add}(\apiname{mv}(\apiname{mul}($\alpha$, A), B), \apiname{mul}($\beta$, C))}.
\end{equation*}
}

The idioms in the \benchmarkname{gemv} kernel are readily apparent, but this is not the case for all benchmarks.
Consider \benchmarkname{doitgen}:

{
\footnotesize
\begin{align*}
&\texttt{build~N~(\debruijnlambda{build~N~(\debruijnlambda{build~N~\debruijnlambda(}}} \\
&\hspace{1.5em}\texttt{ifold~N~0~(\debruijnlambda{\debruijnlambda{A[\debruijnindex{4}][\debruijnindex{3}][\debruijnindex{1}] * B[\debruijnindex{2}][\debruijnindex{1}] + \debruijnindex{0}}}))))}.
\end{align*}
}

\Techniquename finds a surprisingly insightful PyTorch solution:

{
\footnotesize
\begin{equation*}
\irbuildf{N}{\texttt{\apiname{mm}(A[\debruijnindex{0}], \apiname{transpose}(B))}}.
\end{equation*}
}

The \ac{BLAS} \apiname{gemm} function is even trickier to find in \benchmarkname{doitgen}.
\techniquename nonetheless uncovers the idiom by inserting constants and by building a zero matrix using \apiname{memset}.

{
\footnotesize
\begin{align*}
\texttt{build}~\texttt{N}~\texttt{(\debruijnlambda{\gemm{F}{T}(}}&\texttt{1, A[\debruijnindex{0}], B,} \\
&\texttt{1, \irbuildf{N}{\apiname{memset}(0)})}
\end{align*}
}

\techniquename does not find an optimal result for all kernels.
The \benchmarkname{2mm} kernel, for example, could be implemented as \apiname{gemm} calls.
\Techniquename does not find that solution because it would require more saturation steps than the time budget allows for.
Nonetheless, as we will see later, there is still a large speedup obtained by using the detected idioms and the corresponding library calls.

\begin{table}
    \centering
    \footnotesize
    \caption{
        Solutions found for kernels when targeting \ac{BLAS}.
        \textit{Steps} describes the number of saturation steps. %
        \textit{Solution} describes the library calls found at the last step.
        \textit{\eNodes} counts the unique e-nodes in the \egraph, also at the last step.
    }
    \label{tab:benchmark-overview-blas}
    \begin{tabular}{l >{\raggedright}p{3.5cm} r r r}%
        \hline
        \textbf{Kernel} & \textbf{Solution}  & \textbf{Steps} & \bfseries \eNodes \\ %
        \hline
        \begin{filecontents*}{data/benchmark-overview-blas.csv}
name,suite,externs,steps,classes,nodes,time,axpy,dot,gemm,gemv,memset,transpose
2mm,polybench,$3\times\textrm{axpy}$ +~$1\times\textrm{dot}$ +~$1\times\textrm{gemv}$ +~$3\times\textrm{memset}$ +~$1\times\textrm{transpose}$,6,3450,34578,241.680958252,3,1,,1,3,1
atax,polybench,$2\times\textrm{gemv}$ +~$2\times\textrm{memset}$,7,3522,39539,355.413640522,,,,2,2,
doitgen,polybench,$1\times\textrm{gemm}$ +~$1\times\textrm{memset}$,8,2253,47001,318.900629405,,,1,,1,
gemm,polybench,$1\times\textrm{axpy}$ +~$1\times\textrm{gemv}$ +~$1\times\textrm{memset}$,7,3543,49468,420.37769895,1,,,1,1,
gemver,polybench,$2\times\textrm{axpy}$ +~$2\times\textrm{dot}$,5,3071,16923,141.26765801,2,2,,,,
gesummv,polybench,$2\times\textrm{gemv}$ +~$1\times\textrm{memset}$,7,3379,42718,394.388399318,,,,2,1,
jacobi1d,polybench,$1\times\textrm{gemv}$ +~$1\times\textrm{memset}$,5,1844,25330,100.538465315,,,,1,1,
mvt,polybench,$2\times\textrm{gemv}$ +~$2\times\textrm{memset}$,7,2156,26877,166.765805195,,,,2,2,
1mm,custom,$1\times\textrm{gemm}$ +~$1\times\textrm{memset}$,8,2524,44718,366.858512534,,,1,,1,
axpy,custom,$1\times\textrm{axpy}$,11,815,13644,104.334501171,1,,,,,
blur1d,custom,$1\times\textrm{gemv}$ +~$1\times\textrm{memset}$,6,2707,53931,314.9072852,,,,1,1,
gemv,custom,$1\times\textrm{gemv}$,7,2635,34334,241.015227612,,,,1,,
memset,custom,$1\times\textrm{memset}$,11,286,5313,16.238596425,,,,,1,
slim-2mm,custom,$1\times\textrm{gemm}$ +~$1\times\textrm{gemv}$ +~$2\times\textrm{memset}$,7,4030,51764,497.620247079,,,1,1,2,
stencil2d,custom,$1\times\textrm{gemv}$ +~$1\times\textrm{memset}$,5,2127,58834,270.629696287,,,,1,1,
vsum,custom,$1\times\textrm{dot}$,10,1010,15891,73.341919349,,1,,,,
\end{filecontents*}
\csvreader[head to column names]{data/benchmark-overview-blas.csv}{}%
        {\name & \externs  & \steps & \num[round-mode=figures,round-precision=3,exponent-mode=scientific]{\nodes} \\}%
    \end{tabular}
\end{table}

\begin{table}
    \centering
    \footnotesize
    \caption{
        Solutions found for kernels when targeting PyTorch.
        Columns have the same meaning as in \cref{tab:benchmark-overview-blas}.
    }
    \label{tab:benchmark-overview-pytorch}
    \begin{tabular}{l >{\raggedright}p{3.5cm} r r r}%
        \hline
        \textbf{Kernel} & \textbf{Solution}  & \textbf{Steps} & \bfseries \eNodes \\ %
        \hline
        \begin{filecontents*}{data/benchmark-overview-pytorch.csv}
name,suite,externs,steps,classes,nodes,time,add,dot,full,mm,mul,mv,sum,transpose
2mm,polybench,$1\times\textrm{add}$ +~$2\times\textrm{mul}$ +~$2\times\textrm{mv}$ +~$2\times\textrm{transpose}$,5,3983,22827,198.519825779,1,,,,2,2,,2
atax,polybench,$2\times\textrm{mv}$ +~$1\times\textrm{transpose}$,7,3915,19753,210.773570676,,,,,,2,,1
doitgen,polybench,$1\times\textrm{mm}$ +~$1\times\textrm{transpose}$,7,3758,27507,309.221057161,,,,1,,,,1
gemm,polybench,$1\times\textrm{add}$ +~$1\times\textrm{mm}$ +~$2\times\textrm{mul}$,6,5168,26780,454.320588315,1,,,1,2,,,
gemver,polybench,$2\times\textrm{add}$ +~$1\times\textrm{dot}$ +~$1\times\textrm{mul}$ +~$1\times\textrm{mv}$,5,4777,23799,216.978490746,2,1,,,1,1,,
gesummv,polybench,$1\times\textrm{add}$ +~$2\times\textrm{mul}$ +~$2\times\textrm{mv}$,7,6234,31560,526.161863635,1,,,,2,2,,
jacobi1d,polybench,$1\times\textrm{full}$ +~$1\times\textrm{mv}$,5,2880,31253,111.315063699,,,1,,,1,,
mvt,polybench,$2\times\textrm{mv}$ +~$1\times\textrm{transpose}$,7,3409,16875,166.811392899,,,,,,2,,1
1mm,custom,$1\times\textrm{mm}$,7,3541,19853,212.744494584,,,,1,,,,
axpy,custom,$1\times\textrm{add}$ +~$1\times\textrm{mul}$,10,4293,22744,599.930545406,1,,,,1,,,
blur1d,custom,$1\times\textrm{full}$ +~$1\times\textrm{mv}$,5,2067,21272,68.809121835,,,1,,,1,,
gemv,custom,$1\times\textrm{add}$ +~$2\times\textrm{mul}$ +~$1\times\textrm{mv}$,7,5174,26311,367.712609705,1,,,,2,1,,
memset,custom,$1\times\textrm{full}$,11,369,2025,12.503065712,,,1,,,,,
slim-2mm,custom,$1\times\textrm{mm}$ +~$1\times\textrm{mv}$ +~$1\times\textrm{transpose}$,6,3777,20347,226.887670069,,,,1,,1,,1
stencil2d,custom,$1\times\textrm{full}$ +~$1\times\textrm{mv}$,5,4076,90624,509.627878041,,,1,,,1,,
vsum,custom,$1\times\textrm{sum}$,10,3607,17852,238.963047978,,,,,,,1,
\end{filecontents*}
\csvreader[head to column names]{data/benchmark-overview-pytorch.csv}{}%
        {\name & \externs & \steps & \num[round-mode=figures,round-precision=3,exponent-mode=scientific]{\nodes} \\}%
    \end{tabular}
\end{table}

\begin{figure*}
    \centering
    \begin{subfigure}{0.47\textwidth}
        \begin{tikzpicture}
        \begin{axis}[
            axis y line*=left,
        	xlabel=Saturation steps,
            ymax=35000,
        	ylabel=\eNodes,
        	width=0.9\columnwidth,height=4.5cm,
            legend style={at={(0.0,.91)},anchor=west},
            xtick=data
            ]
        \addplot[mark=none, color=blue] coordinates {
    (0, 41)
    (1, 242)
    (2, 614)
    (3, 1892)
    (4, 4320)
    (5, 13411)
    (6, 34334)
}; \label{fig:nodes-over-time-gemm-blas:node-count-plot}
\draw [<-] (axis cs:1,242)-- +(-5pt,8pt) node[left] {\scriptsize $1 \times \texttt{dot}$};
\draw [<-] (axis cs:2,614)-- +(-5pt,11pt) node[left] {\scriptsize $1 \times \texttt{dot}$};
\draw [<-] (axis cs:3,1892)-- +(-5pt,14pt) node[left] {\scriptsize $1 \times \texttt{axpy}, 1 \times \texttt{dot}$};
\draw [<-] (axis cs:4,4320)-- +(-3pt,16pt) node[left] {\scriptsize $2 \times \texttt{axpy}, 1 \times \texttt{dot}, 1 \times \texttt{memset}$};
\draw [<-] (axis cs:5,13411)-- +(-5pt,10pt) node[left] {\scriptsize $2 \times \texttt{axpy}, 1 \times \texttt{dot}, 1 \times \texttt{memset}$};
\draw [<-] (axis cs:6,34334)-- +(-15pt,-5pt) node[left] {\scriptsize $1 \times \texttt{gemv}$};
        \end{axis}

        \begin{axis}[
          axis y line*=right,
          axis x line=none,
          ymax=300,
          width=0.9\columnwidth,height=4.5cm,
          legend style={at={(0.02,0.96)},anchor=north west},
          ylabel={Time [s]},
          xtick=data
        ]

        \addlegendimage{/pgfplots/refstyle=fig:nodes-over-time-gemm-blas:node-count-plot}\addlegendentry{\eNode count}
        \input{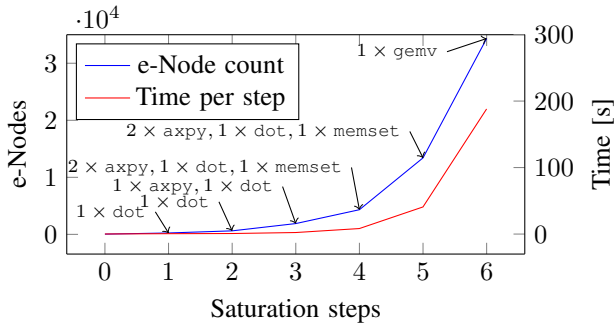}
        \addlegendentry{Time per step}
        \end{axis}
        
        \end{tikzpicture}
    
        \caption{
            Solutions over time for \benchmarkname{gemv}, targeting \ac{BLAS}.
        }
        \label{fig:nodes-over-time-gemm-blas}
    \end{subfigure}%
    \hfill
    \begin{subfigure}{0.47\textwidth}
        \begin{tikzpicture}
        \begin{axis}[
            axis y line*=left,
        	xlabel=Saturation steps,
        	ylabel=\eNodes,
            ymax=35000,
        	width=0.9\columnwidth,height=4.5cm,
            legend style={at={(0.0,.91)},anchor=west},
            xtick=data
            ]
        \addplot[color=blue, mark=none] coordinates {
    (0, 41)
    (1, 273)
    (2, 837)
    (3, 2693)
    (4, 6939)
    (5, 12862)
    (6, 26311)
}; \label{fig:nodes-over-time-gemm-pytorch:node-count-plot}
\draw [<-] (axis cs:1,273)-- +(-5pt,8pt) node[left] {\scriptsize $1 \times \texttt{dot}$};
\draw [<-] (axis cs:2,837)-- +(-5pt,10pt) node[left] {\scriptsize $1 \times \texttt{dot}$};
\draw [<-] (axis cs:3,2693)-- +(5pt,12pt) node[left] {\scriptsize $1 \times \texttt{add}, 1 \times \texttt{dot}, 2 \times \texttt{mul}$};
\draw [<-] (axis cs:4,6939)-- +(-5pt,10pt) node[left] {\scriptsize $1 \times \texttt{add}, 1 \times \texttt{dot}, 2 \times \texttt{mul}$};
\draw [<-] (axis cs:5,12862)-- +(-15pt,5pt) node[left] {\scriptsize $1 \times \texttt{add}, 2 \times \texttt{mul}, 1 \times \texttt{mv}$};
        \end{axis}

        \begin{axis}[
          axis y line*=right,
          axis x line=none,
          ymax=300,
          width=0.9\columnwidth,height=4.5cm,
          legend style={at={(0.02,0.96)},anchor=north west},
          ylabel={Time [s]},
          xtick=data
        ]

        \addlegendimage{/pgfplots/refstyle=fig:nodes-over-time-gemm-blas:node-count-plot}\addlegendentry{\eNode count}
        \input{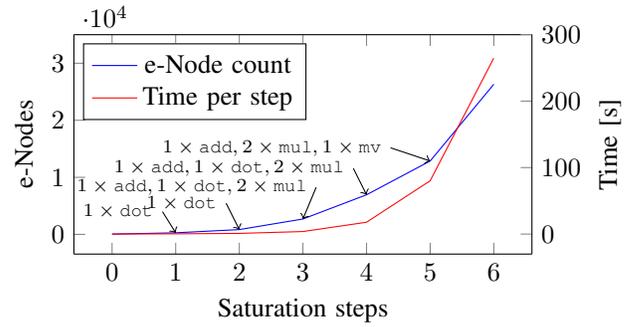}
        \addlegendentry{Time per step}
        \end{axis}
        \end{tikzpicture}
    
        \caption{
            Solutions over time for \benchmarkname{gemv}, targeting PyTorch.
        }
        \label{fig:nodes-over-time-gemv-pytorch}
    \end{subfigure}
    \caption{
        Solutions over time.
        The x-axes show equality saturation steps; the y-axes correspond to the number of \enodes in the \egraph and the amount of time spent performing a saturation step.
        Labeled arrows indicate a new best solution has been found. %
    }
    \label{fig:nodes-over-time}
\end{figure*}

\subsection{Idioms Over Time}
\label{sec:evaluation:idioms-over-time}

The evaluation continues by investigating how \techniquename's solutions evolve.
This evolution is made apparent by plotting kernel solutions over time, as \cref{fig:nodes-over-time} does for the \benchmarkname{gemv} kernel.

\Cref{fig:nodes-over-time-gemm-blas} demonstrates that \techniquename progressively discovers more suitable solutions with each saturation step.
Initially, these solutions consist of dot products.
From steps three to five, \techniquename incorporates vector addition and scaling using \apiname{axpy}.
In step six, these function calls converge into a \apiname{gemv} call.
A similar evolution is observed for PyTorch, as illustrated in \Cref{fig:nodes-over-time-gemv-pytorch}.

\subsection{Coverage}
\label{sec:evaluation:coverage}

\begin{figure}
    \centering
    \begin{tikzpicture}
    \begin{axis}[
        ybar stacked,
        width=\columnwidth,
        height=4cm,
        bar width=15pt,
        ymin=0,
        ymax=1.1,
        legend style={at={(0.03,0.97)}, anchor=north west},
        xlabel=Saturation steps,
        ylabel={Coverage},
        xtick=data,
        ]
        \input{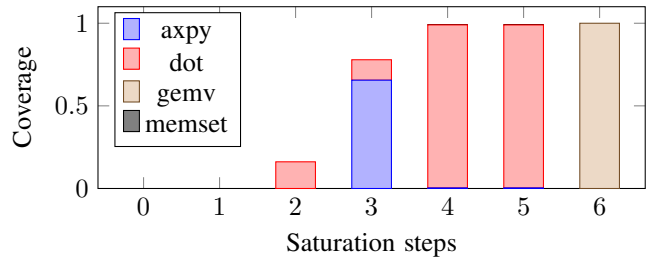}
    \end{axis}
    \end{tikzpicture}
    \caption{
        Coverage over time for the \benchmarkname{gemv} kernel, targeting \ac{BLAS}.
        The x-axis shows saturation steps; the stacked bars depict the ratio of time spent in library functions.
        Higher is better.
    }
    \label{fig:coverage-over-time-gemv-blas}
\end{figure}

We measure the ratio of time kernels spend in the library function to validate \techniquename's effective work offloading.
\Cref{fig:coverage-over-time-gemv-blas} presents the coverage for the \benchmarkname{gemv} kernel, which targets \ac{BLAS}.
The initial \apiname{dot}-based solutions show poor coverage, with negligible coverage at step one and only \SI{16}{\percent} at step two.
However, step three demonstrates a significant improvement, with \apiname{axpy} achieving \SI{66}{\percent} coverage and an additional \SI{12}{\percent} coverage from \apiname{dot}.
Steps four and five favor \apiname{dot}, reaching \SI{99}{\percent} coverage.
Finally, at step six, \apiname{gemv} achieves complete \SI{100}{\percent} coverage.
Although \apiname{memset} appears in two solutions, its coverage contribution is insignificant.

\subsection{Run Time Performance}
\label{sec:evaluation:run-time-performance}

We now examine how \techniquename's idiom recognition affects run-time performance.
For every kernel and every saturation step, we compile \techniquename's pure C and \ac{BLAS} solutions to C code.
We run each solution as many times as we can over the course of one minute and calculate the mean run time.

\begin{figure}
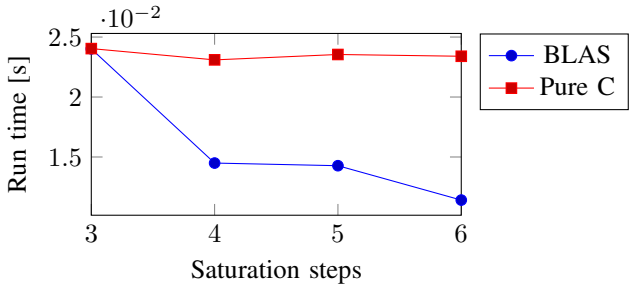

    \centering
    \begin{tikzpicture}
    \begin{axis}[
        xlabel=Saturation steps,
        ylabel={Run time [s]},
        xmin=3,
        xmax=6,
        width=6.5cm,
        height=4cm,
        legend style={at={(1.05,1.0)},anchor=north west},
        restrict y to domain=0.001:0.1,
        ]
        \input{plots/run-time-gemv-blas}
        \input{plots/run-time-gemv-none}
        \legend{\ac{BLAS}, Pure C} %
    \end{axis}
    \end{tikzpicture}
    \caption{
        \benchmarkname{gemv} run times.
        The x-axis shows saturation steps; the y-axis shows the run time of solutions.
        Lower is better.
    }
    \label{fig:run-time-gemv}
\end{figure}

\Cref{fig:run-time-gemv} reports run times for the \benchmarkname{gemv} kernel when targeting \ac{BLAS} and pure C.
Numbers are reported for steps three through six only because the high-level kernel needs to be optimized for a few steps by equality saturation before it can be executed in reasonable time.
Once this is achieved at step three, the pure C and \ac{BLAS} solutions are equally fast.
They diverge as the \ac{BLAS} solution achieves increasingly high coverage.

\begin{figure*}
    \centering
    \begin{tikzpicture}
    \begin{axis}[
        ybar,
        ymode=log,
        width=0.9\textwidth,
        height=4.5cm,
        ymax=2e3,
        ymin=0.5e-3,
        ytick={1e-3,1e-2,1e-1,1,1e1,1e2,1e3},
        ymajorgrids = true,
        bar width=7pt,
        visualization depends on=y\as\myy,
        nodes near coords,
        nodes near coords style = {anchor ={180+sign(\myy)*90},font = \tiny\itshape},
        point meta=rawy,
        ylabel={Speedup},
        enlargelimits=true,
        xticklabel style={rotate=45},
        xtick=data,
        legend style={at={(1.01,1.0)},anchor=north west,nodes={scale=0.8, transform shape}},
        symbolic x coords={2mm,atax,doitgen,gemm,gesummv,jacobi1d,mvt,1mm,axpy,blur1d,gemv,memset,slim-2mm,stencil2d,vsum,dummy,geomean}
        ]
        \input{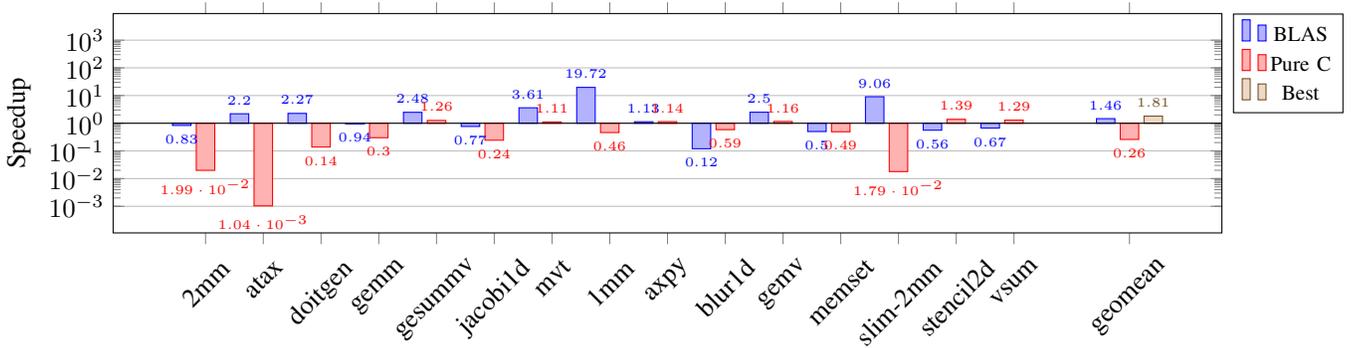}
        \draw (axis cs:{[normalized]\pgfkeysvalueof{/pgfplots/xmin}},1)
            -- (axis cs:{[normalized]\pgfkeysvalueof{/pgfplots/xmax}},1);
    \end{axis}
    \end{tikzpicture}
    \caption{
        Run time speedup of \techniquename's solutions compared to reference implementations in C.
        For PolyBench kernels, the reference implementations are the original benchmarks; for custom benchmarks, they are hand-written C programs coded in the style of PolyBench kernels.
        Each bar represents the quotient of the reference run time and the \techniquename solution run time.
        Higher is better.
    }
    \label{fig:run-time-overview}
\end{figure*}

\Cref{fig:run-time-overview} shows that recognizing idioms results in a run-time speedup of \num{2.5} on \benchmarkname{gemv}.
Speedups vary by kernel.
Notable outliers include \benchmarkname{1mm}, \benchmarkname{vsum}, \benchmarkname{blur1d}, and \benchmarkname{stencil2d}.
The \benchmarkname{1mm} kernel benefits from an ideal \ac{BLAS} solution, netting it a speedup of \num{19.72}.
The cost model replaces the \benchmarkname{vsum} kernel with a \apiname{dot} call, but the associated input array construction outweighs the benefit.
Similarly, the cost model chooses to reduce the convolutions in \benchmarkname{blur1d} and \benchmarkname{stencil2d} to matrix-vector products, performing an im2col transformation.
In practice, this transformation is slower than a direct solution.
The \benchmarkname{gemver} kernel is excluded from the chart as none of its solutions completed within the one-minute time limit.
The geometric mean speedup across all kernels except \benchmarkname{gemver} is \num{1.46} with idiom recognition.
Pure C incurs a slowdown of \num{0.26} on average.
If we choose the fastest solution for each kernel, \techniquename's generated code is \SI{81}{\percent} faster than reference.

\section{Related Work}
\label{sec:related}

There exist a number of works that relate to \techniquename.
\Techniquename's \ac{IR} lies at the intersection of work on the \texttt{build}-\texttt{ifold} paradigm and advances in equality saturation.

\paragraph{\texttt{build}-\texttt{ifold}}

The \texttt{build} and \texttt{ifold} operators were originally introduced in work on bringing \ac{DPS} to functional languages~\cite{shaikhha2017destination}.
In that work, \texttt{build} and \texttt{ifold} were chosen for their similarity to C \texttt{for} loops.
Despite the simplicity of these two operators, further work showed that they can model map-reduce's plethora of data processing primitives~\cite{lin2022from}.
To the best of our knowledge, \techniquename is the first to combine \texttt{build}-\texttt{ifold} with equality saturation.

\paragraph{Equality Saturation}

Another core dependency on which this paper relies is equality saturation.
Originally conceived as a means to optimize Java programs~\cite{tate2009equality}, %
a recent boom in equality saturation research has applied the technique to various specialized problems~\cite{nandi2020synthesizing,vanhattum2021vectorization,wang2020spores,yang2021equality}. %
These problems include the linear algebra domain that forms \techniquename's use case in this paper.
Specifically, one work focuses on sum-product optimization~\cite{wang2020spores} while another optimizes tensor graphs~\cite{yang2021equality}.
\Techniquename differs from these previous studies since it is not domain-specific.
It is a general technique for detecting array processing idioms in a functional \ac{IR}.

\paragraph{\lCalculus and Equality Saturation}

The mentioned use cases avoided name bindings found in \lcalculus.
An initial attempt to encode \lcalculus in \egraphs was demonstrated in a broader work on \texttt{egg}, an optimized equality saturation implementation~\cite{willsey2021egg}.
Another study, which did not involve idiom recognition, improved on \texttt{egg}'s encoding by using De Bruijn indices~\cite{koehler2021sketch}.
It was also the first to examine equality saturation for functional array programming.
\Techniquename adopts both innovations and drastically reduces the number of rewrite rules compared to that prior work by relying on \texttt{build}-\texttt{ifold}, paving the way for a novel approach to idiom recognition.

\paragraph{Idiom Recognition}

Another body of related work is idiom recognition itself.
Idiom rewriting has long served to optimize both imperative~\cite{pinter1994program} and functional languages~\cite{jones2001playing}.
State-of-the-art idiom rewriting research relies on flexible patterns that capture idioms and their variations.

One approach to make patterns more flexible is to encode them as graphs and to find candidate matches using topological embedding~\cite{kawahito2013idiom}.
Dedicated transformations can then massage some of these candidates to match the original pattern.

Patterns can also be explicitly made flexible.
A recent study describes linear algebra patterns in LLVM \ac{IR} using a language called \ac{IDL}~\cite{ginsbach2018automatic}.
The language relies on composable patterns that abstract away variation points such as multiplications by one. %
This allows \ac{IDL} to recognize but not automatically rewrite patterns including BLAS calls.
\textsc{KernelFaRer}, another recent work, takes a similar approach~\cite{carvalho2021kernelfarer}.
It captures variations using flexible idioms that are expressed as LLVM pattern matching components.

By contrast, \ac{SMR}~\cite{espindola2023source} and the MLIR PDL dialect~\cite{mlir2023pdl} describe patterns as source language snippets, making the patterns easier to express at the cost of making them vulnerable to program variations.

\Techniquename stands out from previous approaches by recognizing program variations, as \ac{IDL} and \textsc{KernelFaRer} do, while relying on simple rules in the same language as input programs, in the style of \ac{SMR} and PDL.
\Techniquename accomplishes this feat through its fusion of a minimalist \texttt{build}-\texttt{ifold} \ac{IR} and equality saturation.
The cost of that accomplishment is speed: the approach outlined in this paper is orders of magnitude slower than direct pattern matching, limiting its use for larger kernels.
Future work includes addressing that limitation.

\section{Conclusion}
\label{sec:conclusion}

This paper has addressed idiom recognition for functional array programs by applying equality saturation to a minimalist \ac{IR} with a small set of core rewrite rules.
As seen, this minimalist approach is robust and is able to find idioms that are not explicitly present in the original input program.

Using the \ac{BLAS} and PyTorch libraries as libraries, this paper has shown how the idioms founds in these libraries can be expressed easily using the minimalist \ac{IR}.
The evaluation also demonstrated the robustness of this approach and that it is possible to find  idioms corresponding to these libraries on a set of computational kernels, leading to improved performance.

\section*{Acknowledgements}

We thank Christof Schlaak for implementing the core SHIR project that we build on.
We also thank Zhitao Lin for the SHIR C code back-end that we use to generate C code from high-level kernels.
We acknowledge the support of the Natural Sciences and Engineering Research Council of Canada (NSERC) Discovery Grants Program [grant RGPIN-2020-05889], and the Canada CIFAR AI Chairs Program.
This work was also supported by a Fonds de Recherche du Qu\'{e}bec -- Nature et Technologies 3rd cycle scholarship, award \#304858.

\appendix

\section{Artifact Appendix}

\subsection{Abstract}

This artifact contains a source tree and a Dockerfile.
Docker can assemble these components into a container that includes the \techniquename implementation and its dependencies.
The container is designed to reproduce this paper's experimental results.

\subsection{Artifact check-list (meta-information)}

{\small
\begin{itemize}
  \item {\bf Algorithm: } An equality saturation--based idiom recognition and rewriting algorithm for a minimalist functional array \ac{IR}.

  \item {\bf Compilation: } Docker builds and loads the artifact.
  We used Docker version 20.10.21.

  \item {\bf Data set: } PolyBench/C 4.2.1-beta kernels and custom kernels, included in the artifact.

  \item {\bf Hardware: } A system with one or more x86-64 CPUs.
  We used a server with two 18-core Intel Xeon Gold 6254 CPUs.

  \item {\bf Metrics: } Library calls found, saturation steps, \enode counts, kernel execution times.

  \item {\bf Output: } Kernel optimization and execution logs, graphs and tables generated from log data.

  \item {\bf Experiments: } Kernel optimization with and without idiom recognition (\cref{sec:evaluation:idioms-recognized}, \cref{sec:evaluation:idioms-over-time}); and
  optimized kernel execution (\cref{sec:evaluation:coverage},  \cref{sec:evaluation:run-time-performance}).

  \item {\bf How much time is needed to complete experiments (approximately)?: } Approximately 12--18 hours, varies depending on CPU.

  \item {\bf Publicly available?: } Yes

  \item {\bf Code licenses?: } MIT license

  \item {\bf Archived (provide DOI)?: } 10.5281/zenodo.8316752

\end{itemize}
}

\subsection{Description}

\subsubsection{How delivered}

The artifact is available both on Zenodo (DOI: 10.5281/zenodo.8316752) and in the \texttt{cgo24-artifact} branch of the \texttt{cdubach/shir} BitBucket repository.
Clone it as follows:
\begin{minted}{text}
$ git clone --recursive -b cgo24-artifact \
    https://bitbucket.org/cdubach/shir.git
\end{minted}

\subsubsection{Hardware dependencies} An x86-64 CPU.

\subsubsection{Software dependencies} A Docker installation.

\subsection{Installation}

Build and run a Docker container from the artifact.
In the artifact directory, run the following commands:

\begin{minted}{text}
$ docker build -t liar-image .
$ docker run --name liar-ubuntu -i -t liar-image bash
\end{minted}

\subsection{Experiment workflow}

To replicate our findings, we recommend a workflow consisting of three steps.

\subsubsection{Time-limited optimization}

The first step is to optimize the evaluation's kernels using the same methodology as described in \cref{sec:evaluation}.
That methodology is to optimize each kernel for five minutes for each target (pure C, BLAS idioms or PyTorch idioms).
This experiment will take four hours and its results vary based on single-thread CPU performance.
Faster CPUs will be able to perform more saturation steps and find more advanced solutions.

To perform this experiment, run the following commands in the running Docker container:
\begin{minted}{text}
$ mkdir unlimited-steps
$ cd unlimited-steps
$ ../artifact/src/main/drivers/evaluate_all.py \
    -t300 --optimize-only
$ cd ..
\end{minted}

\subsubsection{Step-limited optimization}

Since time-limited optimization delivers CPU-dependent results, we recommend a step-limited optimization phase to more precisely replicate the findings from \cref{tab:benchmark-overview-blas} and \cref{tab:benchmark-overview-pytorch}.
This second step performs the same optimization process as before, but in a way that produces CPU-invariant results at the cost of requiring a CPU-dependent amount of time.
This trade-off results from a saturation step limit instead of a time limit.
The baked-in step limits are chosen to correspond to the steps reported in the tables.

To perform this second step of the evaluation, run the following:
\begin{minted}{text}
$ mkdir limited-steps
$ cd limited-steps
$ ../artifact/src/main/drivers/evaluate_all.py \
    -t3600 --limit-steps --optimize-only
$ cd ..
\end{minted}

\subsubsection{Kernel execution}

The third step of the workflow involves running the optimized kernels.
We recommend starting from the step-limited results, since these match the solutions reported in \cref{sec:evaluation}.
The following commands will run each solution for one minute, requiring a total of 4\textonehalf{} hours:
\begin{minted}{text}
$ cd limited-steps
$ ../artifact/src/main/drivers/evaluate_all.py \
    -t3600 --limit-steps --build-paper
$ cd ..
\end{minted}

This invocation of the evaluation script will report that it is reusing the optimized kernels from the previous step and then run each solution for each kernel.
Once each solution has been run, the script regenerates this paper from the updated results, allowing for easy inspection of the affected tables and figures.

\subsection{Evaluation and expected result}

Running the evaluation workflow described in the previous section will fill the \texttt{unlimited-steps} and \texttt{limited-steps} directories with subdirectories containing the following:
\begin{enumerate}
    \item Logs for each target, stored in the \texttt{blas-logs}, \texttt{pytorch-logs} and \texttt{none-logs} directories.
    These directories respectively correspond to the BLAS, PyTorch and pure C targets.
    Each log contains detailed information for every equality saturation step.
    Log files whose names have a \texttt{cov-} prefix are derived from their \texttt{log-} equivalents by augmenting them with the results of kernel execution. 

    \item Aggregate data and plots, stored in the \texttt{plots} directory.
\end{enumerate}

The tables and figures in this table are derived from the files in the latter directory.
\begin{itemize}
    \item \Cref{tab:benchmark-overview-blas} is obtained by arranging into a table the following columns from \texttt{blas-overview.csv}: \texttt{name}, \texttt{externs}, \texttt{steps}, and \texttt{nodes}.

    \item \Cref{tab:benchmark-overview-pytorch} is similarly obtained from the data in \texttt{pytorch-overview.csv}.

    \item \Cref{fig:nodes-over-time} is derived by including in a \TeX{} PGFPlots line chart the commands from \path{{nodes,comp-time}-over-time-gemv-{blas,pytorch}.tex}.

    \item \Cref{fig:coverage-over-time-gemv-blas} is a PGFPlots bar chart constructed from the commands in \path{coverage-over-time-gemv-blas.tex}.

    \item \Cref{fig:run-time-gemv} is constructed by creating a PGFPlots line chart from the coordinates in \path{run-time-gemv-blas.tex}, cropped to solutions three through six.

    \item \Cref{fig:run-time-overview} is a PGFPlots bar chart visualization of the commands in \path{speedup-bars.tex}.
\end{itemize}

These tables and figures are generated afresh from locally-computed results by invoking the evaluation script with \texttt{--build-paper}.
The experiment workflow makes use of this flag and stores the resulting paper as \path{paper/conference\_101719.pdf} in the \texttt{limited-steps} directory.

\subsection{Experiment customization}

Experiments can be restricted to specific kernels by passing the names of the kernels to the evaluation script, like so:
\begin{minted}{text}
$ ../artifact/src/main/drivers/evaluate_all.py \
    mvt -t3600 --limit-steps
\end{minted}

\subsection{Notes}

A known discrepancy between the evaluation as performed in \cref{sec:evaluation} and the artifact's Docker container is that the container relies on Ubuntu whereas \cref{sec:evaluation} runs directly on a CentOS system.
We do not believe this difference significantly affects results.

\printbibliography

@String{Computing = "Computing" }

@inproceedings{henriksen17futhark,
author = {Henriksen, Troels and Serup, Niels G. W. and Elsman, Martin and Henglein, Fritz and Oancea, Cosmin E.},
title = {Futhark: Purely Functional GPU-Programming with Nested Parallelism and in-Place Array Updates},
year = {2017},
isbn = {9781450349888},
publisher = {Association for Computing Machinery},
address = {New York, NY, USA},
url = {https://doi.org/10.1145/3062341.3062354},
doi = {10.1145/3062341.3062354},
booktitle = {Proceedings of the 38th ACM SIGPLAN Conference on Programming Language Design and Implementation},
pages = {556–571},
numpages = {16},
location = {Barcelona, Spain},
series = {PLDI 2017}
}

@inproceedings{shaikhha2017destination,
author = {Shaikhha, Amir and Fitzgibbon, Andrew and Peyton Jones, Simon and Vytiniotis, Dimitrios},
title = {Destination-Passing Style for Efficient Memory Management},
year = {2017},
isbn = {9781450351812},
publisher = {Association for Computing Machinery},
address = {New York, NY, USA},
url = {https://doi.org/10.1145/3122948.3122949},
doi = {10.1145/3122948.3122949},
booktitle = {Proceedings of the 6th ACM SIGPLAN International Workshop on Functional High-Performance Computing},
pages = {12–23},
numpages = {12},
location = {Oxford, UK},
series = {FHPC 2017}
}

@inproceedings{lin2022from,
author = {Lin, Zhitao and Dubach, Christophe},
title = {From Functional to Imperative: Combining Destination-Passing Style and Views},
year = {2022},
isbn = {9781450392693},
publisher = {Association for Computing Machinery},
address = {New York, NY, USA},
url = {https://doi.org/10.1145/3520306.3534502},
doi = {10.1145/3520306.3534502},
booktitle = {Proceedings of the 8th ACM SIGPLAN International Workshop on Libraries, Languages and Compilers for Array Programming},
pages = {25–36},
numpages = {12},
location = {San Diego, CA, USA},
series = {ARRAY 2022}
}

@article{willsey2021egg,
author = {Willsey, Max and Nandi, Chandrakana and Wang, Yisu Remy and Flatt, Oliver and Tatlock, Zachary and Panchekha, Pavel},
title = {\texttt{egg}: Fast and Extensible Equality Saturation},
year = {2021},
issue_date = {January 2021},
publisher = {Association for Computing Machinery},
address = {New York, NY, USA},
volume = {5},
number = {POPL},
url = {https://doi.org/10.1145/3434304},
doi = {10.1145/3434304},
journal = {Proc. ACM Program. Lang.},
month = {1},
articleno = {23},
numpages = {29}
}

@article{koehler2021sketch,
  title={Sketch-Guided Equality Saturation: Scaling Equality Saturation to Complex Optimizations in Languages with Bindings},
  author={Koehler, Thomas and Trinder, Phil and Steuwer, Michel},
  journal={arXiv preprint arXiv:2111.13040},
  year={2021}
}

@article{carvalho2021kernelfarer,
author = {De Carvalho, Jo\~{a}o P. L. and Kuzma, Braedy and Korostelev, Ivan and Amaral, Jos\'{e} Nelson and Barton, Christopher and Moreira, Jos\'{e} and Araujo, Guido},
title = {KernelFaRer: Replacing Native-Code Idioms with High-Performance Library Calls},
year = {2021},
issue_date = {September 2021},
publisher = {Association for Computing Machinery},
address = {New York, NY, USA},
volume = {18},
number = {3},
issn = {1544-3566},
url = {https://doi.org/10.1145/3459010},
doi = {10.1145/3459010},
journal = {ACM Trans. Archit. Code Optim.},
month = {6},
articleno = {38},
numpages = {22}
}

@inproceedings{ginsbach2018automatic,
author = {Ginsbach, Philip and Remmelg, Toomas and Steuwer, Michel and Bodin, Bruno and Dubach, Christophe and O'Boyle, Michael F. P.},
title = {Automatic Matching of Legacy Code to Heterogeneous APIs: An Idiomatic Approach},
year = {2018},
isbn = {9781450349116},
publisher = {Association for Computing Machinery},
address = {New York, NY, USA},
url = {https://doi.org/10.1145/3173162.3173182},
doi = {10.1145/3173162.3173182},
booktitle = {Proceedings of the Twenty-Third International Conference on Architectural Support for Programming Languages and Operating Systems},
pages = {139–153},
numpages = {15},
location = {Williamsburg, VA, USA},
series = {ASPLOS '18}
}

@inproceedings{ragankelley2013halide,
author = {Ragan-Kelley, Jonathan and Barnes, Connelly and Adams, Andrew and Paris, Sylvain and Durand, Fr\'{e}do and Amarasinghe, Saman},
title = {Halide: A Language and Compiler for Optimizing Parallelism, Locality, and Recomputation in Image Processing Pipelines},
year = {2013},
isbn = {9781450320146},
publisher = {Association for Computing Machinery},
address = {New York, NY, USA},
url = {https://doi.org/10.1145/2491956.2462176},
doi = {10.1145/2491956.2462176},
booktitle = {Proceedings of the 34th ACM SIGPLAN Conference on Programming Language Design and Implementation},
pages = {519–530},
numpages = {12},
location = {Seattle, Washington, USA},
series = {PLDI '13}
}

@inproceedings{yang2021equality,
 author = {Yang, Yichen and Phothilimthana, Phitchaya and Wang, Yisu and Willsey, Max and Roy, Sudip and Pienaar, Jacques},
 booktitle = {Proceedings of Machine Learning and Systems},
 editor = {A. Smola and A. Dimakis and I. Stoica},
 pages = {255--268},
 title = {Equality Saturation for Tensor Graph Superoptimization},
 url = {https://proceedings.mlsys.org/paper/2021/file/65ded5353c5ee48d0b7d48c591b8f430-Paper.pdf},
 volume = {3},
 year = {2021}
}

@inproceedings{vanhattum2021vectorization,
author = {VanHattum, Alexa and Nigam, Rachit and Lee, Vincent T. and Bornholt, James and Sampson, Adrian},
title = {Vectorization for Digital Signal Processors via Equality Saturation},
year = {2021},
isbn = {9781450383172},
publisher = {Association for Computing Machinery},
address = {New York, NY, USA},
url = {https://doi.org/10.1145/3445814.3446707},
doi = {10.1145/3445814.3446707},
booktitle = {Proceedings of the 26th ACM International Conference on Architectural Support for Programming Languages and Operating Systems},
pages = {874–886},
numpages = {13},
location = {Virtual, USA},
series = {ASPLOS '21}
}

@inproceedings{chandrakana2020synthesizing,
author = {Nandi, Chandrakana and Willsey, Max and Anderson, Adam and Wilcox, James R. and Darulova, Eva and Grossman, Dan and Tatlock, Zachary},
title = {Synthesizing Structured CAD Models with Equality Saturation and Inverse Transformations},
year = {2020},
isbn = {9781450376136},
publisher = {Association for Computing Machinery},
address = {New York, NY, USA},
url = {https://doi.org/10.1145/3385412.3386012},
doi = {10.1145/3385412.3386012},
booktitle = {Proceedings of the 41st ACM SIGPLAN Conference on Programming Language Design and Implementation},
pages = {31–44},
numpages = {14},
location = {London, UK},
series = {PLDI 2020}
}

@inproceedings{tate2009equality,
author = {Tate, Ross and Stepp, Michael and Tatlock, Zachary and Lerner, Sorin},
title = {Equality Saturation: A New Approach to Optimization},
year = {2009},
isbn = {9781605583792},
publisher = {Association for Computing Machinery},
address = {New York, NY, USA},
url = {https://doi.org/10.1145/1480881.1480915},
doi = {10.1145/1480881.1480915},
booktitle = {Proceedings of the 36th Annual ACM SIGPLAN-SIGACT Symposium on Principles of Programming Languages},
pages = {264–276},
numpages = {13},
location = {Savannah, GA, USA},
series = {POPL '09}
}

@INPROCEEDINGS{steuwer2017lift,  author={Steuwer, Michel and Remmelg, Toomas and Dubach, Christophe},  booktitle={2017 IEEE/ACM International Symposium on Code Generation and Optimization (CGO)},   title={LIFT: A functional data-parallel IR for high-performance GPU code generation},   year={2017},  volume={},  number={},  pages={74-85},  doi={10.1109/CGO.2017.7863730}}

@inproceedings{de1972lambda,
  title={Lambda calculus notation with nameless dummies, a tool for automatic formula manipulation, with application to the Church-Rosser theorem},
  author={De Bruijn, Nicolaas Govert},
  booktitle={Indagationes Mathematicae (Proceedings)},
  volume={75},
  number={5},
  pages={381--392},
  year={1972},
  organization={Elsevier}
}

@article{wang2020spores,
  title={SPORES: sum-product optimization via relational equality saturation for large scale linear algebra},
  author={Wang, Yisu Remy and Hutchison, Shana and Leang, Jonathan and Howe, Bill and Suciu, Dan},
  journal={Proceedings of the VLDB Endowment},
  volume={13},
  number={12},
  year={2020}
}

@inproceedings{nandi2020synthesizing,
author = {Nandi, Chandrakana and Willsey, Max and Anderson, Adam and Wilcox, James R. and Darulova, Eva and Grossman, Dan and Tatlock, Zachary},
title = {Synthesizing Structured CAD Models with Equality Saturation and Inverse Transformations},
year = {2020},
isbn = {9781450376136},
publisher = {Association for Computing Machinery},
address = {New York, NY, USA},
url = {https://doi.org/10.1145/3385412.3386012},
doi = {10.1145/3385412.3386012},
booktitle = {Proceedings of the 41st ACM SIGPLAN Conference on Programming Language Design and Implementation},
pages = {31–44},
numpages = {14},
location = {London, UK},
series = {PLDI 2020}
}

@inproceedings{jones2001playing,
  title={Playing by the rules: rewriting as a practical optimisation technique in GHC},
  author={Jones, Simon Peyton and Tolmach, Andrew and Hoare, Tony},
  booktitle={Haskell workshop},
  volume={1},
  pages={203--233},
  year={2001}
}

@article{pinter1994program,
author = {Pinter, Shlomit S. and Pinter, Ron Y.},
title = {Program Optimization and Parallelization Using Idioms},
year = {1994},
issue_date = {May 1994},
publisher = {Association for Computing Machinery},
address = {New York, NY, USA},
volume = {16},
number = {3},
issn = {0164-0925},
url = {https://doi.org/10.1145/177492.177494},
doi = {10.1145/177492.177494},
journal = {ACM Trans. Program. Lang. Syst.},
month = {5},
pages = {305–327},
numpages = {23}
}

@article{blackford2002updated,
author={Blackford, L Susan and Petitet, Antoine and Pozo, Roldan and Remington, Karin and Whaley, R Clint and Demmel, James and Dongarra, Jack and Duff, Iain and Hammarling, Sven and Henry, Greg and others},
title = {An Updated Set of Basic Linear Algebra Subprograms (BLAS)},
year = {2002},
issue_date = {June 2002},
publisher = {Association for Computing Machinery},
address = {New York, NY, USA},
volume = {28},
number = {2},
issn = {0098-3500},
url = {https://doi.org/10.1145/567806.567807},
doi = {10.1145/567806.567807},
journal = {ACM Trans. Math. Softw.},
month = {6},
pages = {135–151},
numpages = {17}
}

@inbook{paszke2019pytorch,
author = {Paszke, Adam and Gross, Sam and Massa, Francisco and Lerer, Adam and Bradbury, James and Chanan, Gregory and Killeen, Trevor and Lin, Zeming and Gimelshein, Natalia and Antiga, Luca and Desmaison, Alban and K\"{o}pf, Andreas and Yang, Edward and DeVito, Zach and Raison, Martin and Tejani, Alykhan and Chilamkurthy, Sasank and Steiner, Benoit and Fang, Lu and Bai, Junjie and Chintala, Soumith},
title = {PyTorch: An Imperative Style, High-Performance Deep Learning Library},
year = {2019},
publisher = {Curran Associates Inc.},
address = {Red Hook, NY, USA},
booktitle = {Proceedings of the 33rd International Conference on Neural Information Processing Systems},
articleno = {721},
numpages = {12}
}

@online{pouchet2016polybench,
author = {Pouchet, Louis-Noël and Yuki, Tomofumi},
title  = {PolyBench/C: the Polyhedral Benchmark suite},
date   = {2016-02-08},
url    = {http://web.cse.ohio-state.edu/~pouchet.2/software/polybench/}
}

@article{espindola2023source,
author = {Espindola, Vinicius and Zago, Luciano and Yviquel, Herv\'{e} and Araujo, Guido},
title = {Source Matching and Rewriting for MLIR Using String-Based Automata},
year = {2023},
issue_date = {June 2023},
publisher = {Association for Computing Machinery},
address = {New York, NY, USA},
volume = {20},
number = {2},
issn = {1544-3566},
url = {https://doi.org/10.1145/3571283},
doi = {10.1145/3571283},
journal = {ACM Trans. Archit. Code Optim.},
month = {3},
articleno = {22},
numpages = {26}
}

@online{mlir2023pdl,
  author = {Riddle, River and B\"{o}ck, Markus and Bandishti, Vinayaka and Bondhugula, Uday and Niu, Jeff and Pienaar, Jacques and Amini, Mehdi},
  title  = {`pdl' Dialect},
  date   = {2023-05-10},
  url    = {https://mlir.llvm.org/docs/Dialects/PDLOps/}
}

@article{kawahito2013idiom,
author = {Kawahito, Motohiro and Komatsu, Hideaki and Moriyama, Takao and Inoue, Hiroshi and Nakatani, Toshio},
title = {Idiom Recognition Framework Using Topological Embedding},
year = {2013},
issue_date = {September 2013},
publisher = {Association for Computing Machinery},
address = {New York, NY, USA},
volume = {10},
number = {3},
issn = {1544-3566},
url = {https://doi.org/10.1145/2512431},
doi = {10.1145/2512431},
journal = {ACM Trans. Archit. Code Optim.},
month = {9},
articleno = {13},
numpages = {34}
}

\end{document}